\documentclass[11pt,fleqn,a4paper]{article}
\usepackage{amssymb,latexsym,amsmath,amsfonts}
\usepackage{graphicx}

    \advance\textheight by \topskip

    \numberwithin{equation}{section}

    \def\tr{{\rm tr \,}}
    \def\Re{{\rm Re \,}}
    \def\Im{{\rm Im \,}}

    \newtheorem{theorem}{Theorem}[section]
    \newtheorem{lemma}[theorem]{Lemma}

    \newtheorem{Definition}[theorem]{Definition}
    
    \newtheorem{Remark}[theorem]{Remark}
    \newenvironment{remark}{\begin{Remark}\rm}{\end{Remark}}
    \newtheorem{Example}[theorem]{Example}

    \newenvironment{proof}%
    {\rm \trivlist \item[\hskip \labelsep{\bf Proof. }]}%
    {\hspace*{\fill}$\Box$\endtrivlist}
    \newenvironment{varproof}%
    {\rm \trivlist \item[\hskip \labelsep{\bf Proof}]}%
    {\hspace*{\fill}$\Box$\endtrivlist}

\begin{document}

    \begin{center} \Large\bf
        Universality for eigenvalue correlations at the origin of the spectrum
    \end{center}

    \

    \begin{center} \large
        A.B.J. Kuijlaars\footnote{Supported by FWO research project G.0176.02 and
            by INTAS project 00-272 and by the Ministry of Science
            and Technology (MCYT) of Spain, project code
            BFM2001-3878-C02-02} \\
            \normalsize \em
            Department of Mathematics, Katholieke Universiteit Leuven, \\
            Celestijnenlaan 200 B,  3001 Leuven, Belgium \\
            \rm arno@wis.kuleuven.ac.be \\[3ex]
            \rm and \\[3ex]
        \large
        M. Vanlessen\footnote{Research Assistant of the Fund for Scientific Research -- Flanders (Belgium)}\\
            \normalsize \em
            Department of Mathematics, Katholieke Universiteit Leuven, \\
            Celestijnenlaan 200 B, 3001 Leuven, Belgium \\
            \rm maarten.vanlessen@wis.kuleuven.ac.be
    \end{center}\ \\[1ex]

\begin{abstract}
    We establish universality of local eigenvalue correlations in
    unitary random matrix ensembles
    $\frac{1}{Z_n}|\det M|^{2\alpha} e^{-n\tr V(M)} dM$
    near the origin of the spectrum.
    If $V$ is even, and if the recurrence
    coefficients of the orthogonal polynomials associated with
    $|x|^{2\alpha} e^{-nV(x)}$ have a regular limiting behavior,
    then it is known from work of Akemann et al., and Kanzieper and
    Freilikher that the local eigenvalue correlations have
    universal behavior described in terms of Bessel functions.
    We extend this to a much wider class of confining
    potentials $V$.
    Our approach is based on the steepest descent method of
    Deift and Zhou for the asymptotic analysis of Riemann-Hilbert
    problems. This method was used by Deift et al.\ to establish
    universality in the bulk of the spectrum. A main part of
    the present work is devoted to the analysis of a
    local Riemann-Hilbert problem near the origin.
\end{abstract}


\section{Introduction}
    \label{section: introduction}

In the present paper we consider the following unitary ensemble of random
matrices, cf.\ \cite{ADMN1,ADMN2}
\begin{equation}\label{unitary-ensemble}
    \frac{1}{Z_n}|\det M|^{2\alpha} e^{-n\tr V(M)} dM,
        \qquad \alpha>-1/2.
\end{equation}
The matrices $M$ are $n\times n$ Hermitian and  $dM$ is the
associated flat Lebesgue measure on the space of
$n \times n$ Hermitian matrices, and $Z_n$ is a normalizing
constant (partition function). The confining potential $V$ in
(\ref{unitary-ensemble}) is a real valued function with enough
increase at infinity, for example a polynomial of even degree with
positive leading coefficient.
Random matrix ensembles are important in many branches of
mathematics and physics, see the recent survey paper
\cite{FSV}. The specific ensemble (\ref{unitary-ensemble}) is
relevant in three-dimensional quantum chromodynamics
\cite{VerbaarschotZahed}.

The ensemble (\ref{unitary-ensemble}) induces a probability density function on
the $n$ eigenvalues $x_1,\ldots, x_n$ of $M$, given by
\[  P^{(n)}(x_1,\ldots ,x_n)=
        \frac{1}{\hat Z_n}\prod_{j=1}^nw_n(x_j)\prod_{i<j}|x_i-x_j|^2,
\]
where $\hat Z_n$ is the normalizing constant, and where $w_n$ is
the following varying weight on the real line,
\begin{equation}\label{varying-weight}
    w_n(x)=|x|^{2\alpha} e^{-nV(x)},
        \qquad\mbox{for $x\in\mathbb{R}$.}
\end{equation}

\medskip

Of particular interest are the local correlations between the
eigenvalues of the ensemble (\ref{unitary-ensemble}) of random
matrices, when their size tends to infinity, at the {\em origin}
of the spectrum, see \cite{ADMN1,KanFrei,Nishigaki}.

The correlations between eigenvalues can be expressed in terms of
the orthonormal polynomials $p_{k,n}(x) = \gamma_{k,n}x^k +
\cdots$ with $\gamma_{k,n} > 0$ with respect to $w_n$, that is
\[
    \int p_{k,n}(x) p_{j,n}(x) |x|^{2\alpha} e^{-nV(x)} dx = \delta_{jk}.
\]
Namely, for $1\leq m\leq n-1$, the $m$-point correlation function
\[
    \mathcal{R}_{n,m}(y_1,\ldots ,y_m)= \frac{n!}{(n-m)!}
        \underbrace{\int_{-\infty}^\infty\ldots\int_{-\infty}^\infty}_{n-m}
        P^{(n)}(y_1,\ldots,y_m,x_{m+1},\ldots ,x_n)dx_{m+1}\ldots d x_n,
\]
satisfies, by a well-known computation of Gaudin and Mehta
\cite{Mehta},
\[
    \mathcal{R}_{n,m}(y_1,\ldots ,y_m)=\det (K_n(y_i,y_j))_{1\leq
    i,j\leq m},
\]
where
\begin{eqnarray}
    \nonumber
    K_n(x,y) &=&
        \sqrt{w_n(x)}\sqrt{w_n(y)}\sum_{j=0}^{n-1}p_{j,n}(x)p_{j,n}(y)
        \\[1ex]
        \label{Kn}
        &=&
        \sqrt{w_n(x)}\sqrt{w_n(y)}\frac{\gamma_{n-1,n}}{\gamma_{n,n}}
        \frac{p_{n,n}(x)p_{n-1,n}(y)-p_{n-1,n}(x)p_{n,n}(y)}{x-y},
\end{eqnarray}
which gives the connection with orthogonal polynomials. The second
equality in (\ref{Kn}) follows from the Christoffel-Darboux
formula \cite{Szego}.

Akemann et al.~\cite{ADMN1} showed
that the local eigenvalue correlations at the origin of the
spectrum have a universal behavior, described in terms of the
following Bessel kernel
\begin{equation}\label{Bessel-kernel}
    \mathbb{J}_{\alpha}^o(u,v)=\pi \sqrt u \sqrt v
    \frac{J_{\alpha+\frac{1}{2}}(\pi u) J_{\alpha-\frac{1}{2}}(\pi v)-
    J_{\alpha-\frac{1}{2}}(\pi u)J_{\alpha+\frac{1}{2}}(\pi
    v)}{2(u-v)},
\end{equation}
where $J_{\alpha\pm\frac{1}{2}}$ denotes the usual Bessel function
of order $\alpha\pm\frac{1}{2}$. In \cite{ADMN1} it was assumed
that the parameter $\alpha$ is a non-negative integer, that the
potential $V$ is even, and that the coefficients $c_{k,n}$ in
the recurrence relation
\[
    xp_{k,n}(x) = c_{k+1,n} p_{k+1,n} + c_{k,n} p_{k-1,n}
\]
satisfied by the orthonormal polynomials have a limiting  behavior
in the sense that the limit $c_{k,n}$ exists whenever $k, n\to
\infty$ such that $k/n \to t$ for some $t>0$. The restriction that
$\alpha$ is a non-negative integer was removed by Kanzieper and
Freilikher \cite{KanFrei}, but they still required the assumption
that $V$ is even and that the recurrence coefficients have a
limiting  behavior. In fact, their method of proof (which they
call Shohat's method) relies heavily on these recurrence
coefficients.

It is the goal of this paper to establish the universality of the
Bessel kernel (\ref{Bessel-kernel}) at the origin of the spectrum without
any assumption on the recurrence coefficients. We can also allow
$V$ to be quite arbitrarily. We assume the following
\begin{eqnarray}
    \label{condition-on-V-eq1}
    \lefteqn{
        \mbox{$V: \mathbb R \to \mathbb R$ is real analytic,}
        } \\[1ex]
    \label{condition-on-V-eq2}
    \lefteqn{
        \lim_{|x|\to\infty}\frac{V(x)}{\log(x^2+1)}=+\infty,
        } \\[1ex]
    \label{condition-on-V-eq3}
    \lefteqn{\psi(0) > 0,}
\end{eqnarray}
where $\psi$ is the density of  the equilibrium measure in the
presence of the external field $V$, \cite{DKM,SaffTotik}.

Let us
explain the condition (\ref{condition-on-V-eq3}). Denote the space
of all probability measures on $\mathbb{R}$ by $M_1(\mathbb{R})$,
and consider the following minimization problem
\begin{equation} \label{minimization}
    \inf_{\mu\in
    M_1(\mathbb{R})}\left(\int\int\log\frac{1}{|s-t|} d\mu(s)d\mu(t)+
    \int V(t)d\mu(t)\right).
\end{equation}
Under the assumptions (\ref{condition-on-V-eq1}) and
(\ref{condition-on-V-eq2}) it is known that the infimum is
achieved \cite{Deift,SaffTotik} uniquely at the
equilibrium measure $\mu_V \in M_1(\mathbb R)$ for $V$.
The measure $\mu_V$ has compact support, and since $V$ is real
analytic, it is supported
on a finite union of intervals. In addition it is absolutely
continuous with respect to the Lebesgue measure, i.e.
\[
    d\mu_V(x)=\psi(x)dx,
\]
and $\psi$ is real analytic on the interior of the support of
$\mu_V$, see \cite{DKM,DKMVZ2}. The importance of the equilibrium
measure lies in the fact that $\psi$ is the limiting (as
$n\to\infty$) mean eigenvalue density of the matrix ensemble
(\ref{unitary-ensemble}), cf.\ \cite{Deift,DKMVZ2}. The condition
(\ref{condition-on-V-eq3}) then says that the mean eigenvalue
density $\psi$ should be strictly positive there. If the origin
belongs to the interior of the support of $\mu_V$ but the mean
eigenvalue density vanishes there, then the potential is called
multicritical, see \cite{ADMN2,AkemannVernizzi,Janik}. This case
will not be treated in this paper.

The regular behavior of the recurrence coefficients assumed
in \cite{ADMN1,KanFrei} is probably satisfied if $V$ is even and if
the support of $\mu_V$ consists of one single interval.
Note that we make no assumptions on the nature of the
support of $\mu_V$. It can consist of any (finite) number
of intervals.

\medskip

Our main result is the following.
\begin{theorem}\label{Theorem: Universality at the origin}
    Assume that the conditions
    {\rm (\ref{condition-on-V-eq1})}--{\rm (\ref{condition-on-V-eq3})} are satisfied.
    Let $w_n$ be the varying weight
    {\rm (\ref{varying-weight})}, let $K_n$ be the
    kernel {\rm (\ref{Kn})}
    associated with $w_n$,  and let $\psi$ be the density of the equilibrium
    measure for $V$.
    Then, for $u,v\in(0,\infty)$,
    \begin{equation}\label{Universality at the origin}
        \frac{1}{n \psi(0)} K_n\left(\frac{u}{n \psi(0)}, \frac{v}{n \psi(0)}\right)
        =
            \mathbb{J}_\alpha^o(u,v)+
            O\left(\frac{u^\alpha v^\alpha }{n}\right),\qquad
            \mbox{as $n\to\infty$,}
    \end{equation}
    where $\mathbb{J}_\alpha^o$ is the Bessel kernel given by
    {\rm (\ref{Bessel-kernel})}.
    The error term in {\rm (\ref{Universality at the origin})}
    is uniform for $u,v$ in bounded subsets of $(0,\infty)$.
\end{theorem}

Other types of universal correlations have been established in the
bulk \cite{BrezinZee,DKMVZ2,KanFrei,PasturShcherbina}, at the soft
edge of the spectrum
\cite{BowickBrezin,Forrester,KanFrei,Moore,TW1}, and at the hard
edge \cite{Forrester,KV,NagaoWadati,TW2}. The universality at the hard
edge is also described in terms of a Bessel kernel, which we have
denoted in \cite{KV} by $\mathbb{J}_\alpha$, namely
\[
    \mathbb{J}_\alpha(u,v)=\frac{J_\alpha(\sqrt u)\sqrt v J'_\alpha(\sqrt v)-
    J_\alpha(\sqrt v)\sqrt u J_\alpha'(\sqrt u)}{2(u-v)}.
\]
To distinguish with this Bessel kernel, we use
$\mathbb{J}_{\alpha}^o$ to denote the Bessel kernel
(\ref{Bessel-kernel}) relevant at the origin of the spectrum.

\begin{remark}
    The universality (\ref{Universality at the origin}) is restricted
    to $u,v >0$. It can be extended to arbitrary real $u$ and $v$
    in the following way. For  $u,v\in\mathbb{R}$, we have that
    \begin{equation} \label{extended universality}
        |u|^{-\alpha}|v|^{-\alpha}\frac{1}{n\psi(0)}
        K_n\left(\frac{u}{n \psi(0)}, \frac{v}{n \psi(0)}\right)
        =
            u^{-\alpha}v^{-\alpha}\mathbb{J}_\alpha^o(u,v)+
            O\left(\frac{1}{n}\right),\qquad
            \mbox{as $n\to\infty$,}
    \end{equation}
    and the error term holds uniformly for $u,v$ in compact
    subsets of $\mathbb{R}$. We will restrict ourselves to
    proving (\ref{Universality at the origin}), but the same
    methods allow us to establish (\ref{extended universality}).
\end{remark}

\medskip

Our proof of Theorem \ref{Theorem: Universality at the origin} is
based on the characterization of the orthogonal polynomials via a
Riemann-Hilbert problem (RH problem) for $2\times 2$ matrix valued
functions, due to Fokas, Its and Kitaev \cite{FokasItsKitaev}, and
on an application of the steepest descent method of Deift and Zhou
\cite{DeiftZhou}. See \cite{Deift,Kuijlaars} for an introduction.
The Riemann-Hilbert approach gives asymptotics for the orthogonal
polynomials in all regions of the complex plane, and it has been
applied before on orthogonal polynomials by a number of authors,
see for example
\cite{BKMM,DKMVZ2,DKMVZ1,KriecherbauerMcLaughlin,KM2,KMVV,Vanlessen}.
Bleher and Its \cite{BleherIts} and Deift et al \cite{DKMVZ2} were
the first to apply Riemann-Hilbert problems  to universality
results in random matrix theory. Later developments include
\cite{BKMM,FyoStrahov,KV}.

In this paper we use many of the ideas of \cite{DKMVZ2}. That
paper deals with the varying weights $e^{-nV(x)}$ with $V$
satisfying (\ref{condition-on-V-eq1}) and
(\ref{condition-on-V-eq2}). The steepest descent method for
Riemann-Hilbert problems is used to establish universality of the
sine kernel in the bulk of the spectrum for the associated unitary
matrix ensembles. In our case the general scheme of the analysis
is the same, and we refer to \cite{DKMVZ2,DKMVZ1} for some of the
details and motivations. The extra factor $|x|^{2\alpha}$
in our weights $|x|^{2\alpha} e^{-nV(x)}$ gives rise to two
important technical differences. The first difference lies in the
construction of the so-called parametrix for the outside region.
To compensate for the factor $|x|^{2\alpha}$ we need to construct
a Szeg\H{o} function on multiple intervals associated to
$|x|^{2\alpha}$. The second and most important difference lies in
the fact that we have to do a local analysis near the origin. This
is where the Bessel functions $J_{\alpha\pm \frac{1}{2}}$ come in.
The construction of the local parametrix near the origin is
analogous to the construction of the parametrix near the algebraic
singularities of the generalized Jacobi weight, recently done by
one of us in \cite{Vanlessen}. The local parametrix determines the
asymptotics of the orthonormal polynomials near the origin, and
thus also governs the universality at the origin of the spectrum.

\medskip

The rest of the paper is organized as follows. In Section
\ref{subsection: associated RH problem} we characterize the
orthogonal polynomials via a RH problem, due to Fokas, Its and
Kitaev \cite{FokasItsKitaev}. Via a series of transformations,
we perform the asymptotic analysis of the RH problem as in
\cite{DKMVZ2,DKMVZ1}. The first transformation will be done in Section
\ref{subsection: first transformation}, the second transformation
in Section \ref{section: second transformation}. Next, we
construct the parametrices for the outside region and near the
origin in Section \ref{section: parametrix for the outside region}
and Section \ref{section: parametrix near the origin},
respectively. The final transformation will be done in Section
\ref{section: third transformation}. Then we have all the
ingredients to prove Theorem \ref{Theorem: Universality at the
origin} in Section \ref{section: proof of the theorem}.  Here we
use some techniques from \cite{KV}.


\section{Associated RH problem and first transformation $Y\mapsto T$}
    \label{section: associated RH problem and first transformation}

In this section we will characterize the orthonormal polynomials
$p_{k,n}$ with respect to the weight (\ref{varying-weight}) as a
solution of a RH problem for a $2\times 2$ matrix valued function
$Y(z)=Y(z;n,w)$, due to Fokas, Its and Kitaev
\cite{FokasItsKitaev}, and do the first transformation in the
asymptotic analysis of this RH problem.

\subsection{Associated RH problem}
    \label{subsection: associated RH problem}

We seek a $2\times 2$ matrix valued function $Y$ that satisfies
the following RH problem.

\subsubsection*{RH problem for \boldmath$Y$:}
\begin{enumerate}
    \item[(a)]
        $Y: \mathbb C \setminus \mathbb R \to \mathbb C^{2\times 2}$
        is  analytic.
    \item[(b)]
        $Y$ possesses continuous boundary values for $x \in \mathbb{R}\setminus\{0\}$
        denoted by $Y_{+}(x)$ and $Y_{-}(x)$, where $Y_{+}(x)$ and $Y_{-}(x)$
        denote the limiting values of $Y(z')$ as $z'$ approaches $x$ from
        above and below, respectively, and
        \begin{equation}\label{RHPYb}
            Y_+(x) = Y_-(x)
            \begin{pmatrix}
                1 & |x|^{2\alpha}e^{-nV(x)} \\
                0 & 1
            \end{pmatrix},
            \qquad\mbox{for $x \in \mathbb{R}\setminus\{0\}$.}
        \end{equation}
    \item[(c)]
        $Y(z)$ has the following asymptotic behavior at infinity:
        \begin{equation} \label{RHPYc}
            Y(z)= \left(I+ O \left( \frac{1}{z} \right)\right)
            \begin{pmatrix}
                z^{n} & 0 \\
                0 & z^{-n}
            \end{pmatrix}, \qquad \mbox{as $z\to\infty$.}
        \end{equation}
    \item[(d)]
        $Y(z)$ has the following behavior near  $z=0$:
        \begin{equation} \label{RHPYd}
            Y(z)=\left\{
            \begin{array}{cl}
                O\begin{pmatrix}
                    1 & |z|^{2\alpha} \\
                    1 & |z|^{2\alpha}
                \end{pmatrix}, &\mbox{if $\alpha< 0$,} \\[2ex]
                O\begin{pmatrix}
                    1 & 1 \\
                    1 & 1
                \end{pmatrix},
                &\mbox{if $\alpha>0$,}
            \end{array}\right.
        \end{equation}
        as $z \to 0$, $z \in \mathbb{C}\setminus\mathbb{R}$.
\end{enumerate}

Compared with the case of no singularity at the origin, see
\cite{DKMVZ2}, we now have an extra condition (\ref{RHPYd}) near
the origin. This condition is used to control the behavior near
the origin, see also \cite{KMVV,Vanlessen}.

\begin{remark}
    The $O$-terms in (\ref{RHPYd}) are to be taken
    entrywise. So for example $Y(z)= O\begin{pmatrix} 1 & |z|^{2\alpha} \\
    1 & |z|^{2\alpha}  \end{pmatrix}$  means that
    $Y_{11}(z) = O(1)$, $Y_{12}(z) = O(|z|^{2\alpha})$, etc.
\end{remark}

The unique solution of the RH problem is given by
\begin{equation}\label{RHPYsolution}
    Y(z) =
    \begin{pmatrix}
        \frac{1}{\gamma_{n,n}}p_{n,n}(z) &
            \frac{1}{2\pi i}\frac{1}{\gamma_{n,n}}
            \int_{-1}^1  \frac{p_{n,n}(x) w_n(x)}{x-z}dx \\[2ex]
        -2\pi i \gamma_{n-1,n}p_{n-1,n}(z) &
            -\gamma_{n-1,n} \int_{-1}^1 \frac{p_{n-1,n}(x)w_n(x)}{x-z} dx
    \end{pmatrix},
\end{equation}
where $p_{k,n}$ is the $k$th degree orthonormal polynomial with
respect to the varying weight $w_n$, and where $\gamma_{k,n}$ is
the leading coefficient of the orthonormal polynomial $p_{k,n}$.
The solution (\ref{RHPYsolution}) is due to Fokas, Its and Kitaev
\cite{FokasItsKitaev}, see also \cite{Deift,DKMVZ2,DKMVZ1}. See
\cite{Kuijlaars,KMVV} for the condition (\ref{RHPYd}).

Note that (\ref{RHPYsolution}) contains the orthonormal
polynomials of degrees $n-1$ and $n$. By (\ref{Kn}) it is then
possible to write $K_n$ in terms of the first column of $Y$. So in
order to prove Theorem \ref{Theorem: Universality at the origin}
an asymptotic analysis of the RH problem for $Y$ is necessary. Via
a series of transformations $Y\to T\to S\to R$ we want to obtain a
RH problem for $R$ which is normalized at infinity (i.e., $R(z)\to
I$ as $z\to\infty$), and with jumps uniformly close to the
identity matrix, as $n\to\infty$. Then $R$ is also uniformly close
to the identity matrix, as $n\to\infty$. Unfolding the series of
transformations, we obtain the asymptotics of $Y$. In particular,
we need the asymptotic behavior of $Y$ near the origin, which
follows from the parametrix near the origin.

\subsection{First transformation $Y\to T$}
    \label{subsection: first transformation}

We first need some properties of the equilibrium measure $\mu_V$
for $V$. Its support is a finite union of disjoint intervals, say
$\bigcup_{j=1}^{N+1} [b_{j-1},a_j]$. So the support consists of
$N+1$ intervals and we refer to these as the {\em bands}. The
complementary $N$ intervals $(a_j,b_j)$ are the {\em gaps}.
Following \cite{DKMVZ2}, we define
\[ J=\bigcup_{j=1}^{N+1}(b_{j-1},a_j) \]
so that $J$ is the interior of the support. The density $\psi$ of
$\mu_V$ has the form  \cite{DKMVZ2}
\begin{equation}\label{definitie-psi}
    \psi(x)=\frac{1}{2\pi i}
    R_+^{1/2}(x)h(x),\qquad\mbox{for $x\in J$,}
\end{equation}
where
\begin{equation} \label{definitie-R}
     R(z)=\prod_{j=1}^{N+1}(z-b_{j-1})(z-a_j),
\end{equation}
and where $h$ is real analytic on $\mathbb{R}$. In this paper we
use $R^{1/2}$ to denote the branch of $\sqrt{R}$ which behaves
like $z^{N+1}$ as $z\to\infty$ and which is defined and analytic
on $\mathbb C \setminus \bar J$. In (\ref{definitie-psi}) we have
that $R_+^{1/2}$ denotes the boundary value of $R^{1/2}$ on $J$
from above. There exists an explicit expression for $h$ in terms of
$V$, see \cite{DKM}, but we will not need that here.

The equilibrium measure minimizes the weighted energy
(\ref{minimization}). The associated  Euler-Lagrange variational
conditions state that there exists a constant $\ell \in\mathbb{R}$
such that
\begin{equation}\label{Variational condition 1}
    2\int\log|x-s|\psi(s)ds-V(x)= \ell,\qquad\mbox{for $x\in \bar J$,}
\end{equation}
\begin{equation}\label{Variational condition 2}
    2\int\log|x-s|\psi(s)ds-V(x)\leq \ell,\qquad\mbox{for $x\in\mathbb{R}\setminus\bar J$.}
\end{equation}
The external field $V$ is called {\em regular} if the
inequality in (\ref{Variational condition 2}) is strict for every
$x \in \mathbb R \setminus \bar J$, and if $h(x)\neq 0$ for every
$x \in \bar J$. Otherwise, $V$ is called {\em singular}. The
regular case holds generically \cite{KM1}. In the singular case
there are a finite number of {\em singular points}. Singular
points in $\bar J$ are such that $h$ vanishes there. Singular
points in $\mathbb R \setminus \bar J$ are such that equality
holds in (\ref{Variational condition 2}).

In order to do the first transformation, we introduce the
so-called $g$-function \cite[Section 3.2]{DKMVZ2}
\begin{equation}\label{definitie-g}
    g(z)=\int \log(z-s)\psi(s)ds,\qquad \mbox{for $z\in\mathbb{C}\setminus(-\infty,a_{N+1}]$,}
\end{equation}
where $\psi(s)ds$ is the equilibrium measure for $V$.
In (\ref{definitie-g}) we take the principal branch of the logarithm,
so that $g$ is analytic on $\mathbb{C}\setminus(-\infty,a_{N+1}]$.

We now give properties of $g$ which are crucial in the following,
\cite[Section 3.2]{DKMVZ2}. From the Euler-Lagrange conditions
(\ref{Variational condition 1}) and (\ref{Variational condition 2})
it follows that
\begin{eqnarray}
    \label{crucial-property-g-1}
    \lefteqn{
        g_+(x)+g_-(x)-V(x)- \ell =0,\qquad\mbox{for $x\in\bar J$,}
        }\\[1ex]
    \label{crucial-property-g-2}
    \lefteqn{
        g_+(x)+g_-(x)-V(x)-\ell \leq 0,
        \qquad\mbox{for $x\in\mathbb{R}\setminus\bar J$.}
        }
\end{eqnarray}
A second crucial property is that
\begin{eqnarray}
    \label{crucial-property-g-3}
    \lefteqn{g_+(x) - g_-(x) = 2\pi i \int_x^{a_{N+1}} d\mu_V(s),
    \qquad\mbox{for $x \in (-\infty, a_{N+1})$,}
    }
\end{eqnarray}
so that $g_+(x)-g_-(x)$ is purely imaginary for all $x \in \mathbb R$
and constant in each of the gaps, namely
\begin{equation}\label{crucial-property-g-4}
    g_+(x)-g_-(x)=
    \left\{\begin{array}{ll}
        2\pi i, &\qquad\mbox{for $x<b_0$,} \\[1ex]
        2\pi i\int_{b_j}^{a_{N+1}} d\mu_V(s)  =: 2\pi i\Omega_j, &\qquad\mbox{for
        $x\in(a_j,b_j),\, j=1\ldots N$,} \\[1ex]
        0, &\qquad\mbox{for $x>a_{N+1}$.}
    \end{array}\right.
\end{equation}
From (\ref{crucial-property-g-4}) we see that $\Omega_j$ is the
total $\mu_V$-mass of the $N+1-j$ largest bands.
These constants all belong to $(0,1)$.
Note that $\Omega_j$ was defined with an extra factor $2\pi$
in \cite{DKMVZ2}.

\medskip
As in \cite[Section 3.3]{DKMVZ2}, we define the matrix valued
function $T$ as
\begin{equation}\label{T-in-function-of-Y}
    T(z)=e^{-\frac{n \ell}{2}\sigma_3}Y(z)e^{\frac{n \ell}{2}\sigma_3}e^{-ng(z)\sigma_3},
        \qquad\mbox{for $z\in\mathbb{C}\setminus\mathbb{R}$,}
\end{equation}
where $\sigma_3= \left(\begin{smallmatrix}
    1 & 0 \\
    0 & -1
\end{smallmatrix}\right)$ is the Pauli matrix. Then $T$ is the
unique solution of the following equivalent RH problem.

\subsubsection*{RH problem for \boldmath$T$:}

\begin{enumerate}
    \item[(a)]
        $T:\mathbb{C}\setminus\mathbb{R}\to \mathbb{C}^{2\times 2}$
        is analytic.
    \item[(b)]
        $T$ satisfies the following jump relations on $\mathbb{R}$:
        \begin{equation}\label{RHPTb1}
            T_{+}(x)=T_{-}(x)
            \begin{pmatrix}
                e^{-n(g_+(x)-g_-(x))} & |x|^{2\alpha} \\
                0 & e^{n(g_+(x)-g_-(x))}
            \end{pmatrix}, \qquad\mbox{for $x\in \bar{J} \setminus\{0\}$,}
        \end{equation}
        \begin{eqnarray}\label{RHPTb2}
            \lefteqn{
                T_+(x)=T_-(x)
                \begin{pmatrix}
                    e^{-2\pi in\Omega_j} & |x|^{2\alpha}e^{n(g_+(x)+g_-(x)-V(x)-\ell)} \\
                    0 & e^{2\pi in\Omega_j}
                \end{pmatrix},}\\[2ex]
            \nonumber
            && \hspace{7,5cm}\mbox{for $x\in(a_j,b_j),j=1\ldots,N$,}
        \end{eqnarray}
        \begin{equation}\label{RHPTb3}
            T_+(x)=T_-(x)
            \begin{pmatrix}
                1 & |x|^{2\alpha} e^{n(g_+(x)+g_-(x)-V(x)-\ell)} \\
                0 & 1
            \end{pmatrix},\quad \mbox{for $x<b_0$ or $x>a_{N+1}$.}
        \end{equation}
    \item[(c)]
        $T(z) = I + O(1/z)$, as $z \to \infty$.
    \item[(d)]
        $T(z)$ has the same behavior as $Y(z)$ as $z\to 0$, given
        by (\ref{RHPYd}).
\end{enumerate}


\section{Second transformation $T\to S$}
    \label{section: second transformation}

In this section we transform the oscillatory diagonal entries of
the jump matrix in (\ref{RHPTb1}) into exponentially decaying
off-diagonal entries. This lies at the heart of the steepest
descent method for RH problems of Deift and Zhou \cite{DeiftZhou},
and this step is often referred to as the opening of the lens.

\medskip

For every $z \in \mathbb C \setminus \mathbb R$ lying
in the region of analyticity of $h$, we define
\[ \phi(z) = \frac{1}{2} \int_z^{a_{N+1}} R^{1/2}(s) h(s) ds \]
where the path of integration does not cross the real axis.
Since $\int_{a_k}^{b_k} R^{1/2}(s) h(s) ds = 0$
for every $k = 1,\ldots, N$, (this follows easily from
the formulas in \cite[Sections 3.1 and 3.2]{DKMVZ2}),
and $\int_{b_j}^{a_{N+1}} R^{1/2}_+(s) h(s) ds = 2\pi i\Omega_j$,
we find that for every $j$,
\begin{eqnarray} \label{phiz upper}
    2\phi(z)&=& \int_{z}^{a_j}R^{1/2}(s)h(s)ds+2\pi i\Omega_j,
        \qquad \mbox{if $\Im z>0$,} \\[1ex]
        \label{phiz lower}
    2\phi(z)&=& \int_{z}^{a_j}R^{1/2}(s)h(s)ds-2\pi i\Omega_j,
        \qquad \mbox{if $\Im z<0$.}
\end{eqnarray}
Note that in \cite{DKMVZ2} a function $G$ is defined  which is
analytic through the bands. We found it more convenient to
have a function with branch cuts along the bands, see also
\cite{Deift}. The functions $G$ and $\phi$ also differ
by a factor $\pm 2$.

The point of the function $\phi$ is that $\phi_+$ and $\phi_-$ are
purely imaginary on the bands, and that
\begin{equation} \label{limits-phi}
    2\phi_+=-2\phi_-=g_+-g_-.
\end{equation}
This means that $2\phi$ and $-2\phi$ provide analytic extensions
of $g_+-g_-$ into the upper half-plane and lower half-plane,
respectively. We also have that for $z$ in a neighborhood of
a regular point $x \in J$, (see \cite[Section 3.3]{DKMVZ2} for
details) that
\begin{equation}\label{ongelijkheid-phi}
    \Re \phi(z)>0, \qquad \mbox{if $\Im z\neq 0$,}
\end{equation}

\medskip

We will now discuss the opening of the lens in the regular case.
In the singular case we need to modify the opening of the lens
somewhat, since we  have to take into account the singular points.
We do not open the lens around  singular points that belong to $J$, see
\cite[Section 4]{DKMVZ2} for details.

For $V$ regular, there is a suitable neighborhood $U$ of $J$ such
that the inequality in (\ref{ongelijkheid-phi}) holds for every
$z\in U$. The opening of the lens is based on the factorization of
the jump matrix (\ref{RHPTb1}) into the following product of three
matrices, see also (\ref{limits-phi}),
\begin{eqnarray}
    \nonumber
    \lefteqn{\begin{pmatrix}
        e^{-n(g_+(x)-g_-(x))} & |x|^{2\alpha} \\
        0 & e^{n(g_+(x)-g_-(x))}
        \end{pmatrix} =
    \begin{pmatrix}
                e^{-2n\phi_+(x)} & |x|^{2\alpha} \\
                0 & e^{-2n\phi_-(x)}
            \end{pmatrix}} \\[1ex]
    && = \begin{pmatrix}
        1 & 0 \\
        |x|^{-2\alpha}e^{-2n\phi_-(x)} & 1
    \end{pmatrix}
    \begin{pmatrix}
        0 & |x|^{2\alpha} \\
        -|x|^{-2\alpha} & 0
    \end{pmatrix}
    \begin{pmatrix}
        1 & 0 \\
        |x|^{-2\alpha}e^{-2n\phi_+(x)} & 1
    \end{pmatrix}.
\end{eqnarray}
As in \cite{Vanlessen} we take an analytic contination of
the factor $|x|^{2\alpha}$ by defining
\begin{equation}\label{definitie-omega}
    \omega(z)=
    \left\{\begin{array}{ll}
        (-z)^{2\alpha},&\qquad\mbox{if $\Re z < 0$,} \\[1ex]
        z^{2\alpha},&\qquad\mbox{if $\Re z > 0$,}
    \end{array}\right.
\end{equation}
with principal branches of powers.
In contrast to the situation in \cite{DKMVZ2}, here we have to
open the lens also going through the origin, cf.\
\cite{Vanlessen}. This follows from the fact that
$|x|^{2\alpha}$ does not have an analytic continuation to a full
neighborhood of the origin.

\begin{figure}
    \center{\resizebox{14cm}{!}{\includegraphics{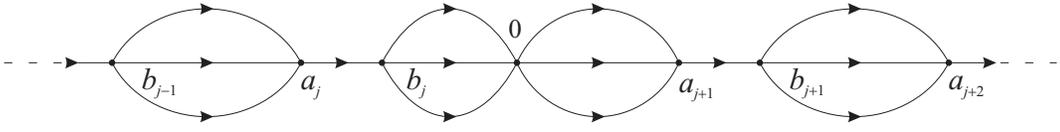}}
    \caption{Part of the contour $\Sigma$.
    }\label{figure: opening-of-the-lens}}
\end{figure}

We thus transform the RH problem for
$T$ into a RH problem for $S$ with jumps on the oriented contour
$\Sigma$, shown in Figure \ref{figure: opening-of-the-lens}. The
precise form of the lens is not yet defined, but it will be
contained in $U$.

Define the piecewise analytic matrix valued function $S$ as
\begin{equation} \label{S-in-function-of-T}
    S(z)=
    \left\{\begin{array}{cl}
        T(z), & \mbox{for $z$ outside the lens,} \\[2ex]
        T(z)
        \begin{pmatrix}
            1 & 0 \\
            -\omega(z)^{-1}e^{-2n\phi(z)} & 1
        \end{pmatrix}, & \mbox{for $z$ in the upper parts of the lens,} \\[2ex]
        T(z)
        \begin{pmatrix}
            1 & 0 \\
            \omega(z)^{-1}e^{-2n\phi(z)} & 1
        \end{pmatrix}, & \mbox{for $z$ in the lower parts of the lens.}
    \end{array} \right.
\end{equation}
Then, $S$ is the unique solution of the following equivalent RH
problem. In (\ref{RHPSb1}), $\mathbb{C}_+$ and $\mathbb{C}_-$ are
used to denote the upper half-plane $\{\Im z>0\}$ and the
lower half-plane $\{\Im z<0\}$, respectively.

\subsubsection*{RH problem for \boldmath$S$:}

\begin{enumerate}
    \item[(a)]
        $S:\mathbb{C}\setminus\Sigma\to\mathbb{C}^{2\times 2}$ is analytic.
    \item[(b)]
        $S$ satisfies the following jump relations on $\Sigma$:
        \begin{equation} \label{RHPSb1}
            S_{+}(z)=
                S_{-}(z)
                \begin{pmatrix}
                    1 & 0 \\
                    \omega(z)^{-1}e^{-2 n\phi(z)} & 1
                \end{pmatrix},
                \qquad\mbox{for $z\in\Sigma\cap\mathbb{C_\pm}$,}
        \end{equation}
        \begin{equation} \label{RHPSb2}
            S_{+}(x)=
                S_{-}(x)
                \begin{pmatrix}
                    0 & |x|^{2\alpha} \\
                    -|x|^{-2\alpha} & 0
                \end{pmatrix},
                \qquad\mbox{for $x\in J\setminus\{0\}$,}
        \end{equation}
        \begin{eqnarray} \label{RHPSb3}
            \lefteqn{
            S_+(x)=
                S_-(x)
                \begin{pmatrix}
                    e^{-2\pi in\Omega_j} & |x|^{2\alpha}e^{n(g_+(x)+g_-(x)-V(x)-\ell)} \\
                    0 & e^{2\pi in\Omega_j}
                \end{pmatrix},} \\[2ex]
            \nonumber
            &&
                \hspace{7,5cm}\mbox{for $x\in(a_j,b_j),j=1\ldots N$,}
        \end{eqnarray}
        \begin{equation} \label{RHPSb4}
            S_+(x) =
                S_-(x)
                \begin{pmatrix}
                    1 & |x|^{2\alpha} e^{n(g_+(x)+g_-(x)-V(x)-\ell)} \\
                    0 & 1
                \end{pmatrix},\quad \mbox{for $x<b_0$ or $x>a_{N+1}$.}
        \end{equation}
    \item[(c)]
        $S(z)=I+O(1/z)$, as $z\to\infty$.
    \item[(d)]
        For $\alpha<0$, the matrix function $S(z)$ has the following behavior as
        $z\to 0$:
        \begin{equation}\label{RHPSd1}
            S(z)=
            O\begin{pmatrix}
                1 & |z|^{2\alpha} \\
                1 & |z|^{2\alpha}
            \end{pmatrix}, \qquad\mbox{as $z\to 0, z\in\mathbb{C}\setminus\Sigma$.}
        \end{equation}
        For $\alpha>0$, the matrix function $S(z)$ has the
        following behavior as $z\to 0$:
        \begin{equation}\label{RHPSd2}
            S(z)=\left\{\begin{array}{cl}
                O\begin{pmatrix}
                    1 & 1 \\
                    1 & 1
                \end{pmatrix},& \mbox{as $z\rightarrow 0$ from outside the lens,} \\[2ex]
                O\begin{pmatrix}
                    |z|^{-2\alpha} & 1 \\
                    |z|^{-2\alpha} & 1
                \end{pmatrix}, & \mbox{as $z\to 0$ from inside the lens.}
            \end{array}\right.
        \end{equation}
    \item[(e)]
    $S$ remains bounded near each of the endpoints $a_i$, $b_j$.
\end{enumerate}

By (\ref{ongelijkheid-phi}) the factor $e^{-2n\phi(z)}$ in
(\ref{RHPSb1}) is exponentially decaying for
$z\in\Sigma\cap\mathbb{C}_\pm$ as $n\to\infty$. This implies that
the jump matrix for $S$ converges exponentially fast to the
identity matrix as $n\to\infty$, on the lips of the lens. Since
$V$ is regular, we have the strict inequality
\begin{equation}\label{strict-equality-regular-case}
    g_+(x)+g_-(x)-V(x)-\ell<0,\qquad \mbox{for
    $x\in\mathbb{R}\setminus\bar J$},
\end{equation}
so that the factor $e^{n(g_+(x)+g_-(x)-V(x)-\ell)}$ in (\ref{RHPSb3})
and (\ref{RHPSb4}) is also exponentially decaying as $n\to\infty$.


\section{Parametrix for the outside region}
    \label{section: parametrix for the outside region}

From the discussion at the end of the previous section we expect
that the leading order asymptotics are determined by the solution
of the following RH problem.

\subsubsection*{RH problem for \boldmath$P^{(\infty)}$:}

    \begin{enumerate}
        \item[(a)]
            $P^{(\infty)}:\mathbb{C}\setminus [b_0,a_{N+1}]\to \mathbb{C}^{2\times 2}$
            is analytic.
        \item[(b)]
            $P^{(\infty)}$ satisfies the following jump relations:
            \begin{equation} \label{RHPPinftyb1}
                P^{(\infty)}_+(x) = P^{(\infty)}_-(x)
                \begin{pmatrix}
                    0 & |x|^{2\alpha} \\
                    -|x|^{-2\alpha} & 0
                \end{pmatrix}, \qquad \mbox{for $x \in J\setminus\{0\}$,}
            \end{equation}
            \begin{equation}\label{RHPPinftyb2}
                P^{(\infty)}_+(x)=P^{(\infty)}_-(x)
                \begin{pmatrix}
                    e^{-2\pi in\Omega_j} & 0 \\
                    0 & e^{2\pi in\Omega_j}
                \end{pmatrix}, \qquad \mbox{for
                $x\in(a_j,b_j),j=1\ldots N$,}
            \end{equation}
        \item[(c)]
            $P^{(\infty)}(z) = I + O\left(1/z \right)$, as $z \to\infty$.
    \end{enumerate}
The solution of this RH problem is referred to as the parametrix
for the outside region, and will be constructed using the
so-called Szeg\H{o} function on the union of disjoint intervals
$J$, associated to $|x|^{2\alpha}$. The importance of the
Szeg\H{o} function is that it transforms this RH problem into a RH
problem with jump matrix $\left(\begin{smallmatrix} 0 & 1 \\ -1 &
0
\end{smallmatrix}\right)$ on $J$.

\subsection{The Szeg\H{o} function}

We seek a scalar function $D : \mathbb C \setminus [b_0, a_{N+1}]
\to \mathbb C$ that solves the following RH problem.

\subsubsection*{RH problem for \boldmath$D$}
\begin{enumerate}
\item[(a)]
    $D$ is non-zero and analytic on $\mathbb{C}\setminus[b_0,a_{N+1}]$
\item[(b)]
    $D$ satisfies the following jump relations:
    \begin{eqnarray}
        \label{RHPDb1}
        \lefteqn{D_+(x)D_-(x)=
            |x|^{2\alpha},\qquad\mbox{for $x\in J\setminus\{0\}$,}
            } \\[1ex]
        \label{RHPDb2}
        \lefteqn{D_+(x)=
            e^{2\pi i\xi_j} D_-(x),
            \qquad\mbox{for $x\in(a_j,b_j),\,j=1,\ldots N$,}
            }
    \end{eqnarray}
    for certain unknown constants $\xi_1,\ldots,\xi_N\in\mathbb{R}$.
    The selection of $\xi_1,\ldots ,\xi_N$ is part of the problem. We
    should choose them such that it is possible to construct $D$.
\item[(c)]
    $D$ and $D^{-1}$ remain bounded near the endpoints
    $a_i,b_j$ of $J$, and
    \begin{equation} \label{definitie Dinfty}
        D_\infty := \lim_{z\to\infty}D(z)
    \end{equation}
    exists and is non-zero.
\end{enumerate}

\medskip

We seek $D$ in the form $D(z)=\exp\Phi(z)$. Then the problem is
reduced  to constructing a scalar function $\Phi$, analytic on
$\mathbb{C}\setminus[b_0,a_{N+1}]$, remaining bounded near the
endpoints $a_i,b_j$ of $J$ and at infinity, and having the following
jumps
\begin{eqnarray}
    \label{jump1-Phi}
    \lefteqn{\Phi_+(x)+\Phi_-(x)=
        2\alpha\log|x|,\qquad \mbox{for $x\in J\setminus\{0\}$,}
        } \\[1ex]
    \label{jump2-Phi}
    \lefteqn{\Phi_+(x)=
        \Phi_-(x)+2\pi i\xi_j,\qquad\mbox{for $x\in(a_j,b_j),\,j=1,\ldots,N$.}
        }
\end{eqnarray}

We can easily check, using Cauchy's formula, the Sokhotskii-Plemelj formula
\cite{Gakhov}, and the fact that $R_-^{1/2}(x)=-R_+^{1/2}(x)$ for $x\in J$,
see (\ref{definitie-R}), that $\Phi$ defined by
\begin{equation}\label{definitie-Phi}
    \Phi(z)=
        R^{1/2}(z)\left(\frac{1}{2\pi i}
        \int_J\frac{2\alpha\log|x|}{R_+^{1/2}(x)}\frac{dx}{x-z}
        + \sum_{j=1}^N \xi_j
        \int_{a_j}^{b_j}\frac{1}{R^{1/2}(x)}\frac{dx}{x-z}\right),
\end{equation}
satisfies the jump conditions (\ref{jump1-Phi}) and (\ref{jump2-Phi}). We
note that $\Phi$ is analytic on $\mathbb{C}\setminus
[b_0,a_{N+1}]$ and remains bounded near the endpoints $a_i,b_j$
of $J$. We  use the freedom we have in choosing the constants
$\xi_1,\ldots, \xi_N$ to ensure that $\Phi$ remains bounded at infinity.
Since $R^{1/2}(z)$ behaves like $z^{N+1}$ as $z\to\infty$, and since
\[
    \frac{1}{x-z}= - \sum_{k=0}^{N-1} \frac{x^k}{z^{k+1}}
    + O\left(\frac{1}{z^{N+1}}\right),
    \qquad \mbox{as $z\to\infty$},
\]
we have to choose $\xi_1,\ldots,\xi_N$ such that the $N$
conditions
\begin{equation}\label{systemequations}
    \frac{1}{2\pi i}\int_J \frac{2\alpha\log|x|}{R_+^{1/2}(x)}\,x^k
    dx+\sum_{j=1}^{N}\xi_j\int_{a_j}^{b_j}\frac{x^k dx}{R^{1/2}(x)}\,=0,\qquad k=0,\ldots ,N-1,
\end{equation}
are satisfied. Note that (\ref{systemequations}) represents a
system of $N$ linear equations with coefficient matrix
\begin{equation}\label{definitie-A}
    A=
    \begin{pmatrix}
        \int_{a_1}^{b_1}\frac{dx}{R^{1/2}(x)} &
            \int_{a_2}^{b_2}\frac{dx}{R^{1/2}(x)} & \cdots &
            \int_{a_N}^{b_N}\frac{dx}{R^{1/2}(x)} \\[2ex]
        \int_{a_1}^{b_1}\frac{xdx}{R^{1/2}(x)} &
            \int_{a_2}^{b_2}\frac{xdx}{R^{1/2}(x)} & \cdots &
            \int_{a_N}^{b_N}\frac{xdx}{R^{1/2}(x)} \\[2ex]
        \vdots & \vdots & \ddots & \vdots \\[2ex]
            \int_{a_1}^{b_1}\frac{x^{N-1}dx}{R^{1/2}(x)} &
            \int_{a_2}^{b_2}\frac{x^{N-1}dx}{R^{1/2}(x)} & \cdots &
            \int_{a_N}^{b_N}\frac{x^{N-1}dx}{R^{1/2}(x)}
    \end{pmatrix}.
\end{equation}
By the multilinearity of the determinant, we have
\begin{eqnarray} \nonumber
    \det A & = &
    \int_{a_1}^{b_1} \ldots \int_{a_N}^{b_N} \det
    \begin{pmatrix}
        1  & 1  & \cdots & 1 \\[2ex]
        x_1 & x_2 & \cdots & x_N \\[2ex]
        \vdots & \vdots & \ddots & \vdots \\[2ex]
        x_1^{N-1} & x_2^{N-1} & \cdots & x_N^{N-1}
    \end{pmatrix}
        \frac{dx_1}{R^{1/2}(x_1)} \cdots \frac{dx_N}{R^{1/2}(x_N)} \\
    & = & \label{determinant-A}
    \int_{a_1}^{b_1} \ldots
    \int_{a_N}^{b_N} \prod_{j<k}(x_k-x_j)\frac{dx_1}{R^{1/2}(x_1)} \cdots
    \frac{dx_N}{R^{1/2}(x_N)}.
\end{eqnarray}
Since the gaps $(a_j,b_j)$ are disjoint, we have $x_j<x_k$
for $j<k$, so that $\prod_{j<k}(x_k-x_j)>0$. Using the fact that
$R^{1/2}$ does not change sign on each of the gaps $(a_j,b_j)$, it
then follows that the integrand in (\ref{determinant-A}) has
a constant sign in the region of integration, so that
$\det A\neq 0$ and $A$ is invertible.

We then define $\xi_1,\ldots ,\xi_N$ as follows
\begin{equation}\label{definitie-xi}
    \begin{pmatrix}
        \xi_1 \\
        \xi_2 \\
        \vdots \\
        \xi_N
    \end{pmatrix}
    = -A^{-1}
    \begin{pmatrix}
        \frac{1}{2\pi i}\int_J \frac{2\alpha\log|x|}{R_+^{1/2}(x)}\,dx \\[2ex]
        \frac{1}{2\pi i}\int_J \frac{2\alpha\log|x|}{R_+^{1/2}(x)}\,xdx \\[2ex]
        \vdots \\[2ex]
        \frac{1}{2\pi i}\int_J \frac{2\alpha\log|x|}{R_+^{1/2}(x)}\,x^{N-1}dx
    \end{pmatrix}.
\end{equation}
Note that  $R^{1/2}$ is real on each of the gaps, so that by
(\ref{definitie-A}) all entries of $A$ are real. This implies,
from (\ref{definitie-xi}) and the fact that $R^{1/2}_+$ is purely
imaginary on $J$, that the constants $\xi_1, \ldots, \xi_N$ are
real.

We proved the following
\begin{theorem} \label{Theorem: Szego function}
    The scalar function $D(z)=\exp\Phi(z)$, where $\Phi$ is
    given by {\rm (\ref{definitie-Phi})}, and the constants
    $\xi_1,\ldots, \xi_N$ by {\rm (\ref{definitie-xi})}, solves the RH
    problem for $D$.
\end{theorem}

For later use we state the following lemma.
\begin{lemma}\label{gedrag-D-rond-0}
    We have that $z^{-\alpha} D(z)$ and $z^{\alpha} D(z)^{-1}$
    remain bounded near the origin.
\end{lemma}

\begin{proof}
    For definiteness, suppose that the origin lies on the band $(b_j,a_{j+1})$ with
    $j\in\{0,\ldots ,N\}$. Since $D(z)=\exp\Phi(z)$, with $\Phi$ given by
    (\ref{definitie-Phi}), it is sufficient to prove that
    \begin{equation}\label{gedrag-D-rond-0-proof}
        \frac{1}{2\pi
        i}\int_{b_j}^{a_{j+1}}\frac{2\alpha\log|x|}{R_+^{1/2}(x)}\frac{dx}{x-z}=
        \frac{\alpha\log z}{R^{1/2}(z)}+F(z),
    \end{equation}
    with $F$ analytic near the origin. Fix $z$ near
    the origin with $\Im z \neq 0$, and let
    $\gamma_\delta$ be the oriented contour shown
    in Figure \ref{figure: contour-lemma}, with $\delta>0$ small. Cauchy's formula
    implies
    \[
        \frac{1}{2\pi i}\int_{\gamma_\delta}\frac{\log\zeta}{R^{1/2}(\zeta)}\frac{d\zeta}{\zeta-z}=
        \frac{\log z}{R^{1/2}(z)}.
    \]
    \begin{figure}
        \center{\resizebox{9cm}{!}{\includegraphics{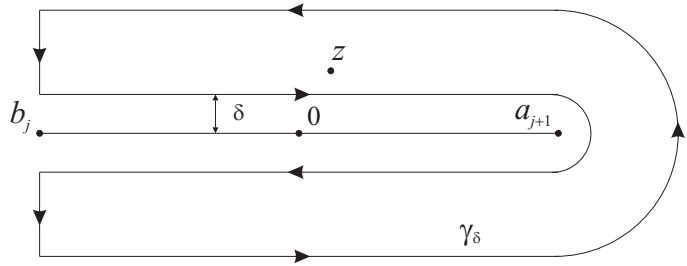}}
        \caption{The oriented contour $\gamma_\delta$.}\label{figure: contour-lemma}}
    \end{figure}
    Letting $\delta\to 0$, we then have, since $R_+^{1/2}(x)=-R_-^{1/2}(x)$
    for $x\in (b_j,a_{j+1})$,
    \begin{eqnarray*}
        \lefteqn{
            \tilde{F}(z)+\frac{1}{2\pi i}\int_{b_j}^0\frac{\log|x|+i\pi}{R_+^{1/2}(x)}\frac{dx}{x-z}+
            \frac{1}{2\pi i}\int_0^{a_{j+1}}\frac{\log|x|}{R_+^{1/2}(x)}\frac{dx}{x-z}
        } \\[2ex]
        &&
        \qquad-\frac{1}{2\pi i}\int_{a_{j+1}}^{0}\frac{\log|x|}{R_+^{1/2}(x)}\frac{dx}{x-z}-
        \frac{1}{2\pi i}\int_{0}^{b_{j}}\frac{\log|x|-i\pi}{R_+^{1/2}(x)}\frac{dx}{x-z}
        = \frac{\log z}{R^{1/2}(z)},
    \end{eqnarray*}
    with $\tilde{F}$ analytic near the origin.
    Hence
    \[
        \frac{1}{2\pi
        i}\int_{b_j}^{a_{j+1}}\frac{\log|x|}{R_+^{1/2}(x)}\frac{dx}{x-z}=\frac{1}{2}\frac{\log
        z}{R^{1/2}(z)}-\frac{1}{2}\tilde{F}(z),
    \]
    so that (\ref{gedrag-D-rond-0-proof}) holds with $F(z) = -\alpha \tilde{F}(z)$,
    which proves the lemma.
\end{proof}

\subsection{Construction of $P^{(\infty)}$}

    We now use the Szeg\H{o} function $D$ from the previous subsection,
    to transform the RH problem for $P^{(\infty)}$ into a RH problem with
    jump matrix $\left(\begin{smallmatrix}0 & 1\\ -1 & 0\end{smallmatrix}\right)$ on
    $J$. We seek $P^{(\infty)}$ in the form, cf.\ \cite{KMVV,Vanlessen}
    \begin{equation}\label{N-in-function-of-N1}
        P^{(\infty)}(z)=D_\infty^{\sigma_3}\tilde P^{(\infty)}(z)D(z)^{-\sigma_3}, \qquad
        \mbox{for $z\in\mathbb{C}\setminus[b_0,a_{N+1}]$.}
    \end{equation}
    Then, by (\ref{RHPPinftyb1})--(\ref{RHPDb2}) the problem is reduced to
    constructing a solution of the
    following RH problem.

    \subsubsection*{RH problem for \boldmath$\tilde P^{(\infty)}$}
    \begin{enumerate}
    \item[(a)]
        $\tilde P^{(\infty)}:\mathbb{C}\setminus [b_0,a_{N+1}]\to\mathbb{C}^{2\times 2}$
        is analytic.
    \item[(b)]
        $\tilde P^{(\infty)}$ satisfies the following jump relations:
        \begin{equation}
            \tilde P^{(\infty)}_+(x)=\tilde P^{(\infty)}_-(x)
            \begin{pmatrix}
                0 & 1 \\
                -1 & 0
            \end{pmatrix},\qquad\mbox{for $x\in J\setminus\{0\}$,}
        \end{equation}
        \begin{eqnarray}\label{RHPN1b2}
            \lefteqn{\tilde P^{(\infty)}_+(x)=\tilde P^{(\infty)}_-(x)
                \begin{pmatrix}
                    e^{-2\pi in\Omega_j}e^{2\pi i\xi_j} & 0 \\
                    0 & e^{2\pi in\Omega_j}e^{-2\pi i\xi_j}
                \end{pmatrix},}\\[2ex]
            \nonumber
            && \hspace{6cm}\mbox{for $x\in (a_j,b_j),\,j=1,\ldots ,N$.}
        \end{eqnarray}
    \item[(c)]
        $\tilde P^{(\infty)}(z)=I+O(1/z)$, as $z\to\infty$.
    \end{enumerate}
    This corresponds to the RH problem \cite[(4.24)--(4.26)]{DKMVZ2},
    which has been solved there using Riemann theta
    functions. Note that, in contrast to the
    RH problem \cite[(4.24)--(4.26)]{DKMVZ2}, the jump matrix in
    (\ref{RHPN1b2}) contains extra factors $\exp(\pm 2\pi i\xi_j)$ in the
    diagonal entries, which come from the Szeg\H{o} function $D$. However,
    this does not create any problems.

    \medskip

    In order to formulate the solution of the RH problem for $\tilde P^{(\infty)}$ we
    need to introduce some additional notations. Here we closely follow
    \cite{DKMVZ2}, see also \cite{DIZ}.

\begin{figure}[h]
        \center{\resizebox{11cm}{!}{\includegraphics{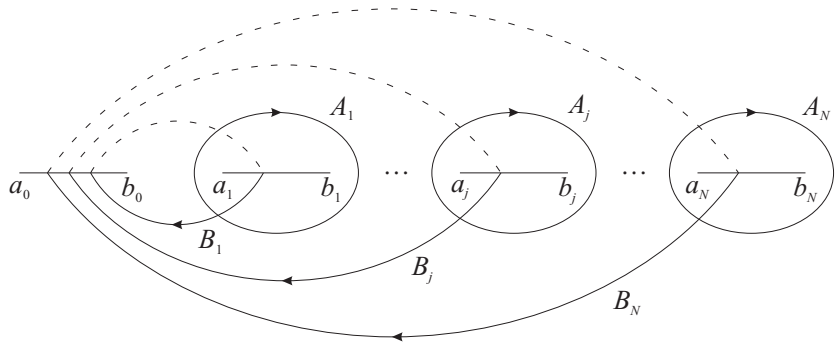}}
        \caption{The canonical homology basis $\{A_i,B_j\}_{1\leq i,j\leq
        N}$ for $X$. The full lines denote paths on the first sheet, while the dotted
        lines denote paths on the second sheet.}\label{figure: canonical-homology-basis}}
    \end{figure}

    Let $\tilde J = \mathbb R \setminus \bar J$  be the
    complement of $\bar J$, and $a_0\equiv a_{N+1}$.
    Letting the point $\infty$ lie on the interval $(a_0,b_0)$,
    $\tilde J$ can be displayed as a union of intervals on the
    Riemann sphere. Let $X$ be the two-sheeted Riemann surface of genus $N$
    associated to $\sqrt{R(z)}$, obtained by gluing together  two copies
    of the slit plane $\mathbb{C}\setminus\tilde J$ along $\tilde
    J$. We draw cycles $A_j$ winding once, in the negative direction, around the
    slit $(a_j,b_j)$ in the first sheet, and cycles $B_j$
    starting from a point on the slit $(a_j,b_j)$ going on the
    first sheet trough a point on the slit $(a_0,b_0)$, and
    returning on the second sheet to the original point, as
    indicated in Figure \ref{figure: canonical-homology-basis}. The cycles $\{A_i,B_j\}_{1\leq i,j\leq
    N}$ form a canonical homology basis for $X$, see
    \cite{FarkasKra}.

    Let $\omega=(\omega_1,\ldots ,\omega_N)$ be the basis of
    holomorphic one-forms on $X$ dual to the canonical homology
    basis, that is
    \begin{equation}\label{normalization-Aj}
        \int_{A_j}\omega_i=\delta_{ij},\qquad 1\leq i,j\leq N.
    \end{equation}
    The associated Riemann matrix of $B$ periods, denoted by
    $\tau$ and with entries
    \begin{equation}
        \tau_{ij}=\int_{B_j}\omega_i,\qquad 1\leq i,j\leq N,
    \end{equation}
    is symmetric with positive definite imaginary part, see
    \cite{FarkasKra}. The associated Riemann theta function is
    defined by
    \begin{equation}
        \theta(z)=\sum_{m\in\mathbb{Z}^N}\exp 2\pi i
        \left(\langle m,z\rangle+\frac{1}{2}\langle m,\tau
        m\rangle\right),\qquad z\in \mathbb{C}^N,
    \end{equation}
    where $\langle\cdot,\cdot\rangle$ is the real scalar product,
    which defines an analytic function on $\mathbb C^N$. The Riemann theta
    function has the periodicity properties \cite{FarkasKra} with
    respect to the lattice $\mathbb{Z}^N+\tau\mathbb{Z}^{N}$
    \begin{equation}\label{periodicity}
        \theta(z+e_j)=\theta(z),\qquad\theta(z\pm\tau_j)=e^{\mp 2\pi i z_j -\pi
        i\tau_{jj}}\theta(z),
    \end{equation}
    where $z = (z_1, \ldots, z_N)$ and $e_j$ is the $j$th unit vector
    in $\mathbb{C}^N$  with $1$ on  the $j$th entry and zeros elsewhere,
    and where $\tau_j$ is the  $j$th column vector of $\tau$.

    Define the scalar function
    \begin{equation}
        \gamma(z)=\left[\prod_{i=1}^N\left(\frac{z-b_i}{z-a_i}\right)
            \left(\frac{z-b_0}{z-a_{N+1}}\right)\right]^{1/4}
    \end{equation}
    which is analytic on $\mathbb{C}\setminus\tilde J$, with $\gamma(z)\sim
    1$ as $z\to\infty,z\in\mathbb{C}_+$. It is known \cite[Lemma 4.1]{DKMVZ2}
    that $\gamma$ has the following properties:
    \[
        \mbox{$\gamma+\gamma^{-1}$ possesses $N$ roots
        $\{z_j^{(-)}\}_{j=1}^N$ with $z_j^{(-)}$ on the $-$ side of
        $(a_j,b_j)$,}
    \]
    \[
        \mbox{$\gamma-\gamma^{-1}$ possesses $N$ roots
        $\{z_j^{(+)}\}_{j=1}^N$ with $z_j^{(+)}$ on the $+$ side of
        $(a_j,b_j)$.}
    \]

    Fix the base point for the Riemann surface $X$ to be
    $a_{N+1} = a_0$, let $K$ be the associated vector of Riemann
    constants \cite{FarkasKra}, and define the multivalued function
    \begin{equation}
        u(z)=\int_{a_{N+1}}^z\omega.
    \end{equation}
    Here, we take the integral along any path from $a_{N+1}$ to
    $z$ on the first sheet. Since the integral is taken on the first
    sheet, $u(z)$ is uniquely defined in
    $\mathbb{C}^N/\mathbb{Z}^N$
    because of (\ref{normalization-Aj}). Let $d$ be defined as
    \begin{equation}
        d=-K-\sum_{j=1}^{N}\int_{a_{N+1}}^{z_j^{(-)}} \omega,
    \end{equation}
    where again the integrals are taken on the first sheet.

    \medskip

    We now have introduced the necessary ingredients to formulate the solution of
    the RH problem for $\tilde P^{(\infty)}$. Together with
    (\ref{N-in-function-of-N1}) this gives the parametrix $P^{(\infty)}$ for the
    outside region. The solution of the RH problem for $\tilde P^{(\infty)}$ is given by, see
    \cite[Lemma 4.3]{DKMVZ2},
    \begin{eqnarray}
        \nonumber
        \tilde P^{(\infty)}(z) &=& {\rm diag}
        \begin{pmatrix}
            \frac{\theta(u_+(\infty)+d)}
            {\theta(u_+(\infty)-n
            \Omega+\xi+d)}, \frac{\theta(u_+(\infty)+d)}
            {\theta(-u_+(\infty)-n \Omega+\xi-d)}
        \end{pmatrix} \\[2ex]
        \label{solution-N1-eq1}
        &&
            \qquad\times\,
            \begin{pmatrix}
                \frac{\gamma+\gamma^{-1}}{2}\,
                    \frac{\theta(u(z)-n\Omega+\xi+d)}{\theta(u(z)+d)} &
                \frac{\gamma-\gamma^{-1}}{-2i}\,
                    \frac{\theta(-u(z)-n\Omega+\xi+d)}{\theta(-u(z)+d)}
                \\[2ex]
                \frac{\gamma-\gamma^{-1}}{2i}\,
                    \frac{\theta(u(z)-n\Omega+\xi-d)}{\theta(u(z)-d)} &
                \frac{\gamma+\gamma^{-1}}{2}\,
                    \frac{\theta(-u(z)-n\Omega+\xi-d)}{\theta(u(z)+d)}
            \end{pmatrix},
    \end{eqnarray}
    for $z\in\mathbb{C}_+$, and
    \begin{eqnarray}
        \nonumber
        \tilde P^{(\infty)}(z) &=& {\rm diag}
        \begin{pmatrix}
            \frac{\theta(u_+(\infty)+d)}
            {\theta(u_+(\infty)-n
            \Omega+\xi+d)}, \frac{\theta(u_+(\infty)+d)}
            {\theta(-u_+(\infty)-n
            \Omega+\xi-d)}
        \end{pmatrix} \\[2ex]
        \label{solution-N1-eq2}
        &&
            \qquad\times\,
            \begin{pmatrix}
                \frac{\gamma-\gamma^{-1}}{-2i}\,
                    \frac{\theta(-u(z)-n\Omega+\xi+d)}{\theta(-u(z)+d)} &
                -\frac{\gamma+\gamma^{-1}}{2}\,
                    \frac{\theta(u(z)-n\Omega+\xi+d)}{\theta(u(z)+d)}
                \\[2ex]
                \frac{\gamma+\gamma^{-1}}{2}\,
                    \frac{\theta(-u(z)-n\Omega+\xi-d)}{\theta(u(z)+d)} &
                -\frac{\gamma-\gamma^{-1}}{2i}\,
                    \frac{\theta(u(z)-n\Omega+\xi-d)}{\theta(u(z)-d)}
            \end{pmatrix},
    \end{eqnarray}
    for $z\in\mathbb{C}_-$. Here,
    $\Omega=(\Omega_1,\ldots, \Omega_N)$ and $\xi=(\xi_1,\ldots, \xi_N)$.

    \begin{remark}
        In contrast to \cite{DKMVZ2} we have an extra term $\xi$ in the
        Riemann theta functions. This comes from the slightly
        different jump matrix in (\ref{RHPPinftyb2}) due to the Szeg\H{o}
        function, as noted before. If $\xi\in\mathbb{Z}^N$
        the factors $e^{\pm 2\pi i\xi_j}$ in (\ref{RHPN1b2}) disappear and
        the RH problem for $\tilde P^{(\infty)}$ is exactly the same as the RH
        problem \cite[(4.24)--(4.26)]{DKMVZ2}. Since the Riemann theta
        functions possess the periodicity properties (\ref{periodicity}), the
        term $\xi$ in
        (\ref{solution-N1-eq1}) and (\ref{solution-N1-eq2}) disappears in this case. This
        is in agreement with \cite[Lemma 4.3]{DKMVZ2}.
    \end{remark}

    For later use, we need $P^{(\infty)}$ to be invertible. In \cite[Section 4.2]{DKMVZ2}
    it has been shown that $\det\tilde P^{(\infty)}\equiv 1$, so
    that by (\ref{N-in-function-of-N1})
    \begin{equation}\label{det-Pinfty=1}
        \det P^{(\infty)}\equiv 1.
    \end{equation}


\section{Parametrix near the origin}
    \label{section: parametrix near the origin}

In this section we construct the parametrix near the origin. As
noted in the introduction, it is similar to the construction of
the parametrix near the algebraic singularities of the generalized
Jacobi weight \cite{Vanlessen}, and we skip some details and
motivations.

We surround the origin by a disk $U_\delta$ with radius $\delta>0$.
We assume that $\delta$ is small, so that in any case, we
have that $[-\delta, \delta] \subset J$.
We seek a matrix valued function $P$ that satisfies the
following RH problem.

\subsubsection*{RH problem for \boldmath$P$:}
\begin{enumerate}
\item[(a)]
    $P(z)$ is defined and analytic for $z \in
    U_{\delta_0}\setminus\Sigma$ for some $\delta_0 > \delta$.
\item[(b)]
    On $\Sigma \cap U_{\delta}$, $P$ satisfies the same jump relations
    as $S$, that is,
    \begin{eqnarray}
        \label{RHPPb1}
        P_+(z) &=&
            P_-(z)
            \begin{pmatrix}
                1 & 0 \\
                \omega(z)^{-1}e^{-2n\phi(z)} & 1
            \end{pmatrix},
            \qquad\mbox{for $z\in \Sigma \cap \left(U_\delta
                \cap\mathbb{C}_\pm\right)$,}
        \\[1ex]
        \label{RHPPb2}
        P_+(x) &=&
            P_-(x)
            \begin{pmatrix}
                0 & |x|^{2\alpha} \\
                -|x|^{-2\alpha} & 0
            \end{pmatrix},
            \qquad \mbox{for $x \in (-\delta, \delta) \setminus\{0\}$.}
    \end{eqnarray}
\item[(c)]
    On $\partial U_\delta$ we have, as $n \to \infty$
    \begin{equation}\label{RHPPc}
        P(z) \left(P^{(\infty)}\right)^{-1}(z) = I + O \left( \frac{1}{n} \right),
        \qquad \mbox{uniformly for $z \in \partial U_\delta\setminus\Sigma$.}
    \end{equation}
\item[(d)]
    For $\alpha<0$, the matrix function $P(z)$ has the following behavior as $z\to 0$:
    \begin{equation}\label{RHPPd1}
        P(z)=
        O\begin{pmatrix}
            1 & |z|^{2\alpha} \\
            1 & |z|^{2\alpha}
        \end{pmatrix},
        \qquad \mbox{as $z\to 0$.}
    \end{equation}
    For $\alpha>0$, the matrix function $P(z)$ has the following behavior as $z\to 0$:
    \begin{equation}\label{RHPPd2}
        P(z)=
        \left\{\begin{array}{cl}
            O\begin{pmatrix}
                1 & 1 \\
                1 & 1
            \end{pmatrix},
            & \mbox{as $z\to 0$ from outside the lens,}\\[3ex]
            O\begin{pmatrix}
                |z|^{-2\alpha} & 1 \\
                |z|^{-2\alpha} & 1
            \end{pmatrix},
            & \mbox{as $z\to 0$ from inside the lens.}
        \end{array}\right.
    \end{equation}
\end{enumerate}

We construct $P$ as follows. First, we focus on conditions (a),
(b) and (d). We transform the RH problem for $P$ into a RH problem
for $P^{(1)}$ with constant jump matrices, and solve the latter RH
problem explicitly. Afterwards, we also consider the matching
condition (c) of the RH problem.

\medskip

We start with the following map $f$ defined on a neighborhood of the
origin
\begin{equation}\label{fx0}
    f(z)=
    \left\{\begin{array}{ll}
        i\phi(z)-i\phi_+(0), & \mbox{if $\Im z>0$,} \\[1ex]
        -i\phi(z)-i\phi_+(0), & \mbox{if $\Im z<0$.}
    \end{array}\right.
\end{equation}
Since $\phi_+ = -\phi_-$, we have that $f$ is analytic
for $z$ in a neighborhood of the origin. An
easy calculation, based on the fact that
$2\phi_+(x)=g_+(x)-g_-(x)$ and on (\ref{definitie-g}) and
(\ref{fx0}), shows that
\begin{equation}\label{f-real}
    f(x)=\pi \int_0^x\psi(s)ds,\qquad
    \mbox{for $x\in (-\delta, \delta)$,}
\end{equation}
which implies that $f'(0)=\pi \psi(0) >0$. So, the behavior of $f$
near the origin is given by
\begin{equation}\label{behavior-f}
    f(z)=\pi\psi(0)z+O\left(z^2\right), \qquad \mbox{as $z\to 0$.}
\end{equation}
So if we choose $\delta>0$ sufficiently
small, $\zeta=f(z)$ is a conformal mapping on $U_\delta$ onto a
convex neighborhood of $0$ in the complex $\zeta$-plane.
We also note that $f(x)$ is real and positive (negative) for
$x \in U_{\delta}$ positive (negative), which follows from (\ref{f-real}).

\begin{figure}
    \center{\resizebox{6cm}{!}{\includegraphics{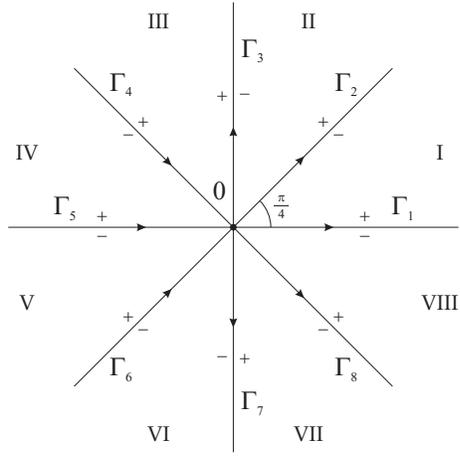}}
    \caption{The contour $\Gamma_{\Psi}$.}\label{figure: system-of-contours-Psi}}
\end{figure}

\begin{figure}
    \center{\resizebox{12cm}{!}{\includegraphics{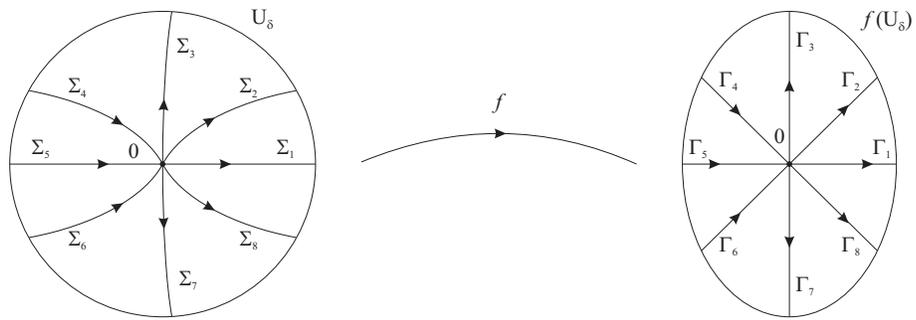}}
    \caption{The conformal mapping $f$. Every $\Sigma_k$ is mapped onto the part
    of the corresponding ray $\Gamma_k$ in $f(U_\delta)$.}
    \label{figure: conformal-mapping-f}}
\end{figure}

Let $\Gamma_j,j=1,\ldots ,8$ be the infinite ray
\[
    \Gamma_j = \{ \zeta \in \mathbb C \mid \arg \zeta = (j-1)\frac{\pi}{4} \}.
\]
These rays divide the $\zeta$-plane into eight sectors I--VIII
as shown in Figure \ref{figure: system-of-contours-Psi}.
We  define the contours
$\Sigma_j$, $j=1,2, \ldots, 8$ as the preimages under the
mapping $\zeta = f(z)$ of the part of the corresponding rays
$\Gamma_j$ in $f(U_{\delta})$,  see Figure \ref{figure: conformal-mapping-f}.

We have some freedom in the selection of the contour
$\Sigma$. We now specify that we open the lens
in such a way that
\[
    \Sigma \cap U_{\delta} = \bigcup\limits_{j=1,2,4,5,6,8} \Sigma_j.
\]
As a consequence we have that $f$ maps $\Sigma$ to part of
the union of rays $\bigcup_j \Gamma_j$.

In order to transform to constant jumps we use a piecewise
analytic function $W$ corresponding to the analytic continuation
of $|x|^{2\alpha}$. For $z \in U_{\delta}$, we define
\begin{equation}\label{definitie-W}
    W(z) =
        \left\{\begin{array}{ll}
            z^\alpha, & \qquad \mbox{if $\pi/2 < |\arg f(z)| < \pi$,} \\[1ex]
            (-z)^\alpha, & \qquad \mbox{if $0 < |\arg f(z)| < \pi/2$,}
        \end{array}\right.
\end{equation}
with principal branches of powers.
Then $W$ is defined and analytic in $U_{\delta} \setminus
(\Sigma_1 \cup \Sigma_3 \cup \Sigma_5 \cup \Sigma_7)$.

We seek $P$ in the form
\begin{equation}\label{P in function of P1}
    P(z)=
        E_n(z)P^{(1)}(z)W(z)^{-\sigma_3}e^{-n\phi(z)\sigma_3}.
\end{equation}
Here the matrix valued function $E_n$ is analytic in a
neighborhood of $U_\delta$, and $E_n$ will be determined below so
that the matching condition (c) of the RH problem for $P$ is
satisfied. Similar considerations as in \cite{Vanlessen} show that
$P^{(1)}$ should satisfy the following RH problem, with jumps on
the system of contours $\bigcup_{i=1}^8\Sigma_i$,
oriented as in the left part of  Figure \ref{figure: conformal-mapping-f}.
In (\ref{RHPP1b1})--(\ref{RHPP1b4}),
$\Sigma_i^o$ is used to denote $\Sigma_i$ without the origin.


\subsubsection*{RH problem for \boldmath$P^{(1)}$:}
\begin{enumerate}
\item[(a)]
    $P^{(1)}(z)$ is defined and analytic for $z\in
    U_{\delta_0}\setminus(\Sigma\cup\Gamma)$ for some $\delta_0 > \delta$.
\item[(b)]
    $P^{(1)}$ satisfies the following jump relations on $U_{\delta}\cap(\Sigma\cup\Gamma)$:
    \begin{eqnarray}
        \label{RHPP1b1}
        P_+^{(1)}(x)
        &=&
            P_-^{(1)}(x)
            \begin{pmatrix}
                0 & 1 \\
                -1 & 0
            \end{pmatrix},
            \qquad \mbox{for $x\in \Sigma_1^o\cup\Sigma_5^o$,} \\[2ex]
        \label{RHPP1b2}
        P_+^{(1)}(z)
        &=&
            P_-^{(1)}(z)
            \begin{pmatrix}
                1 & 0 \\
                e^{-2\pi i\alpha} & 1
            \end{pmatrix},
            \qquad \mbox{for $z\in \Sigma_2^o\cup\Sigma_6^o$,} \\[2ex]
        \label{RHPP1b3}
        P_+^{(1)}(z)
        &=&
            P_-^{(1)}(z) e^{\pi i \alpha\sigma_3},
            \qquad \mbox{for $z\in \Sigma_3^o\cup\Sigma_7^o$,} \\[2ex]
        \label{RHPP1b4}
        P_+^{(1)}(z)
        &=&
            P_-^{(1)}(z)
            \begin{pmatrix}
                1 & 0 \\
                e^{2\pi i\alpha} & 1
            \end{pmatrix},
            \qquad \mbox{for $z\in \Sigma_4^o \cup\Sigma_8^o$.}
    \end{eqnarray}
\item[(c)]
    For $\alpha<0$, $P^{(1)}(z)$ has the following behavior as $z\to 0$:
    \begin{equation}\label{RHPP1c1}
        P^{(1)}(z)=
        O\begin{pmatrix}
            |z|^{\alpha} & |z|^{\alpha} \\
            |z|^{\alpha} & |z|^{\alpha}
        \end{pmatrix},
        \qquad \mbox{as $z\to 0$.}
    \end{equation}
    For $\alpha>0$, $P^{(1)}(z)$ has the following behavior as $z\to 0$:
    \begin{equation}\label{RHPP1c2}
        P^{(1)}(z)=
        \left\{\begin{array}{cl}
            O\begin{pmatrix}
                |z|^{\alpha} & |z|^{-\alpha} \\
                |z|^{\alpha} & |z|^{-\alpha}
            \end{pmatrix},
            & \mbox{as $z\to 0$ from outside the lens,} \\[3ex]
            O\begin{pmatrix}
                |z|^{-\alpha} & |z|^{-\alpha} \\
                |z|^{-\alpha} & |z|^{-\alpha}
            \end{pmatrix},
            & \mbox{as $z\to 0$ from inside the lens.}
        \end{array}\right.
    \end{equation}
\end{enumerate}

\medskip

Next we construct an explicit solution of the RH problem for
$P^{(1)}$. This is based on a model RH problem for $\Psi_\alpha$
in the $\zeta$-plane, see \cite{Vanlessen}. We denote by
$\Gamma_{\Psi}$ the contour $\bigcup_{j=1}^8 \Gamma_j$ oriented as
shown in Figure \ref{figure: system-of-contours-Psi}.

\subsubsection*{RH problem for \boldmath$\Psi_\alpha$:}
\begin{enumerate}
\item[(a)]
    $\Psi_\alpha:\mathbb{C}\setminus\Gamma_{\Psi}\to\mathbb{C}^{2\times 2}$ is analytic.
\item[(b)]
    $\Psi_\alpha$ satisfies the following jump relations on $\Gamma_\Psi$:
    \begin{eqnarray} \label{jumpPsi15}
        \Psi_{\alpha,+}(\zeta)
        &=&
            \Psi_{\alpha,-}(\zeta)
            \begin{pmatrix}
                0 & 1 \\
                -1 & 0
            \end{pmatrix},
            \qquad \mbox{for $\zeta \in \Gamma_1\cup\Gamma_5$,} \\[2ex]
        \Psi_{\alpha,+}(\zeta)
        &=& \label{jumpPsi26}
            \Psi_{\alpha,-}(\zeta)
            \begin{pmatrix}
                1 & 0 \\
                e^{-2\pi i \alpha}  & 1
            \end{pmatrix},
            \qquad \mbox{for $\zeta \in \Gamma_2\cup\Gamma_6$,} \\[2ex]
        \Psi_{\alpha,+}(\zeta)
        &=& \label{jumpPsi37}
            \Psi_{\alpha,-}(\zeta)
            e^{\pi i \alpha\sigma_3},
            \qquad \mbox{for $\zeta \in \Gamma_3\cup\Gamma_7$,} \\[2ex]
        \Psi_{\alpha,+}(\zeta)
        &=& \label{jumpPsi48}
            \Psi_{\alpha,-}(\zeta)
            \begin{pmatrix}
                1 & 0 \\
                e^{2\pi i \alpha}  & 1
            \end{pmatrix},
            \qquad \mbox{for $\zeta \in \Gamma_4\cup\Gamma_8$.}
    \end{eqnarray}
\item[(c)]
    For $\alpha<0$ the matrix function $\Psi_\alpha(\zeta)$ has the following behavior as $\zeta\to 0$:
    \begin{equation}\label{RHPPSIx0c1}
        \Psi_\alpha(\zeta)=
        O\begin{pmatrix}
            |\zeta|^\alpha & |\zeta|^\alpha \\
            |\zeta|^\alpha & |\zeta|^\alpha
        \end{pmatrix},
        \qquad \mbox{as $\zeta\to 0$.}
    \end{equation}
    For $\alpha>0$ the matrix function $\Psi_\alpha(\zeta)$ has the following behavior as $\zeta\to 0$:
    \begin{equation}\label{RHPPSIx0c2}
        \Psi_\alpha(\zeta)=
        \left\{\begin{array}{cl}
            O\begin{pmatrix}
                |\zeta|^\alpha & |\zeta|^{-\alpha} \\
                |\zeta|^\alpha & |\zeta|^{-\alpha}
            \end{pmatrix},
            & \mbox{as $\zeta\to 0$ with $\zeta\in$ II, III, VI, VII,} \\[3ex]
            O\begin{pmatrix}
                |\zeta|^{-\alpha} & |\zeta|^{-\alpha}\\
                |\zeta|^{-\alpha} & |\zeta|^{-\alpha}
            \end{pmatrix},
            & \mbox{as $\zeta\to 0$ with $\zeta\in$ I, IV, V, VIII.}
        \end{array}\right.
    \end{equation}
\end{enumerate}
This RH problem was solved in  \cite[formulas
(4.26)--(4.33)]{Vanlessen}. It is built out of the modified Bessel
functions $I_{\alpha\pm \frac{1}{2}}$, $K_{\alpha\pm\frac{1}{2}}$
and out of the Hankel functions $H^{(1)}_{\alpha\pm\frac{1}{2}}$,
$H^{(2)}_{\alpha\pm\frac{1}{2}}$. For our purpose here, it
suffices to know the explicit formula for $\Psi_\alpha$ in sector
I. There we have
\begin{equation}\label{PsiAlphaI}
    \Psi_\alpha(\zeta)=\frac{1}{2}\sqrt\pi\zeta^{1/2}
    \begin{pmatrix}
        H_{\alpha+\frac{1}{2}}^{(2)}(\zeta) &
            -iH_{\alpha+\frac{1}{2}}^{(1)}(\zeta) \\[2ex]
        H_{\alpha-\frac{1}{2}}^{(2)}(\zeta) &
            -iH_{\alpha-\frac{1}{2}}^{(1)}(\zeta)
    \end{pmatrix}e^{-(\alpha+\frac{1}{4})\pi
    i\sigma_3},\qquad\mbox{for $0<\arg\zeta<\frac{\pi}{4}$.}
\end{equation}
Starting from (\ref{PsiAlphaI}) we can find the solution in the other
sectors by following the jumps (\ref{jumpPsi15})--(\ref{jumpPsi48}).
See \cite{Vanlessen} for explicit expressions.

Now we define
\begin{equation}
    P^{(1)}(z)=\Psi_\alpha(nf(z)),
\end{equation}
and $P^{(1)}$ will solve the RH problem for $P^{(1)}$. This ends
the construction of $P^{(1)}$.

\medskip

So far, we have proven that for every matrix valued function $E_n$
analytic in a neighborhood of $U_\delta$, the matrix valued
function $P$ given by
\begin{equation}\label{definitie-P}
    P(z) =
    E_n(z)\Psi_\alpha(nf(z))W(z)^{-\sigma_3}e^{-n\phi(z)\sigma_3},
\end{equation}
satisfies conditions (a), (b) and (d) of the RH problem for $P$.
We now use the freedom we have in choosing $E_n$ to ensure that
$P$, given by (\ref{definitie-P}), also satisfies the matching
condition (c) of the RH problem for $P$. To this end, we use the
asymptotic behavior of $\Psi_\alpha$ at infinity, see
\cite[(4.43)--(4.46)]{Vanlessen}. Similar calculations as in
\cite{Vanlessen} show that we have to define $E_n$ as
\begin{equation}\label{Enx0}
    E_n(z)= E(z) e^{n\phi_+(0)\sigma_3} e^{-\frac{\pi i}{4}\sigma_3} \frac{1}{\sqrt 2}
    \begin{pmatrix}
        1 & i \\
        i & 1
    \end{pmatrix},
\end{equation}
where the matrix valued function $E$ is given by
\begin{eqnarray}
    \label{NTildeQuadrant1}
    E(z) &=&
        P^{(\infty)}(z) W(z)^{\sigma_3} e^{\frac{1}{2}\alpha\pi
        i\sigma_3}, \qquad\mbox{for $z\in f^{-1}(I \cup II)$,} \\[2ex]
    \label{NTildeQuadrant2}
    E(z) &=&
        P^{(\infty)}(z) W(z)^{\sigma_3} e^{-\frac{1}{2}\alpha\pi
        i\sigma_3}, \qquad\mbox{for $z\in f^{-1}(III \cup IV)$,} \\[2ex]
    \label{NTildeQuadrant3}
    E(z) &=&
        P^{(\infty)}(z) W(z)^{\sigma_3}
        \begin{pmatrix}
            0 & 1 \\
            -1 & 0
        \end{pmatrix}
        e^{-\frac{1}{2}\pi i \alpha\sigma_3},
        \qquad\mbox{for $z\in f^{-1}(V \cup VI)$,} \\[2ex]
    \label{NTildeQuadrant4}
    E(z)
    &=&
        P^{(\infty)}(z) W(z)^{\sigma_3}
        \begin{pmatrix}
            0 & 1 \\
            -1 & 0
        \end{pmatrix}
        e^{\frac{1}{2}\pi i \alpha\sigma_3},  \qquad\mbox{for $z\in f^{-1}(VII \cup VIII)$.}
\end{eqnarray}

Following the proof of \cite[Proposition 4.5]{Vanlessen}, we
obtain that $E$ is analytic in a full neighborhood of
$U_{\delta}$. Here we need the fact that $D(z)/W(z)$ and
$W(z)/D(z)$ remain bounded as $z \to 0$, which follows from
(\ref{definitie-W}) and Lemma \ref{gedrag-D-rond-0}. Then we see
from (\ref{Enx0}) that $E_n$ is also analytic in a neighborhood of
$U_{\delta}$. This completes the construction of the parametrix
near the origin.

\begin{remark}
    Note that, in contrast to the case of the generalized Jacobi weight \cite{Vanlessen},
    here $E$ depends on $n$. This follows from the fact that the
    parametrix $P^{(\infty)}$ for the outside region in our case depends on
    $n$.
\end{remark}

For later use we state, since $E$ is analytic in $U_\delta$ and
from the explicit form of $P^{(\infty)}$, cf.\ \cite{DKMVZ2}, that
$E(z)$ and $\frac{d}{dz}E(z)$ are uniformly bounded for $z\in
U_\delta$, as $n\to\infty$.


\section{Third transformation $S\to R$}
    \label{section: third transformation}

At each of the endpoints $a_i$, $b_j$ of $J$, we have to do a
local analysis as well as at each of the singular points (if any).
The endpoints and singular points are surrounded by small disks,
say of radius $\delta$, that do not overlap and that also do
not overlap with the disk $U_{\delta}$ around the region.
Within each disk we construct a parametrix $P$
which satisfies a local RH problem:

\subsubsection*{RH problem for \boldmath$P$ near $x_0$ where $x_0$
is an endpoint or a singular point:}
\begin{enumerate}
\item[(a)]
    $P(z)$ is defined and analytic for $z \in
    \{ |z-x_0| < \delta \}\setminus\Sigma$ for some $\delta_0 > \delta$.
\item[(b)]
    $P$ satisfies the same jump relations as $S$ does on
    $\Sigma \cap \{|z-x_0| < \delta\}$.
\item[(c)]
    There is $\kappa > 0$ such that we  have
    as $n \to \infty$:
    \begin{equation}\label{RHPPcbis}
        P(z) \left(P^{(\infty)}\right)^{-1}(z) = I +
        O \left( \frac{1}{n^{\kappa}} \right),
        \qquad \mbox{uniformly for $|z-x_0| = \delta$.}
    \end{equation}
\end{enumerate}

The local RH problem near the regular endpoints $a_i$, $b_j$ of $J$ is
similar to the situation in \cite{DKMVZ2}. Here however, we have
extra factors $|x|^{\pm 2\alpha}$ and $\omega(z)^{-1}$ in the jump
matrices. These factors can easily be removed  via an
appropriate transformation, and the local RH problem is then
solved as in \cite[Section 4.3--Section 4.5]{DKMVZ2} with the
use of  Airy functions. For our purpose, we do not need the
explicit formulas
for the parametrix near the endpoints. It suffices to know that $P$
exists. For regular endpoints we can take $\kappa = 1$
in  (\ref{RHPPcbis}).

Near the singular points we can follow the analysis
of \cite[Section 5]{DKMVZ2}. Here we do not construct an
explicit parametrix out of special functions, only existence
of the local RH problem is obtained. For singular points
we have $\kappa < 1$ in (\ref{RHPPcbis}), see \cite{DKMVZ2}.
So the singular points
lead to error   terms with decay slower than for the regular
points $(\kappa=1)$. However, this has no influence on the
universality result at the origin (not even in the error term),
since that only depends on the leading order asymptotics.

\medskip

We are now ready to do the final transformation. As noted before,
we surround the endpoints $a_i,b_j$ of $J$, the origin, and the
singular points of the potential $V$ by nonoverlapping small
disks.  Using the parametrix $P^{(\infty)}$ for the
outside region and the parametrix $P$ defined inside each of the disks,
we define the matrix valued function $R$ as
\begin{equation}\label{R-in-function-of-S}
    R(z)=\left\{
    \begin{array}{ll}
        S(z)\left(P^{(\infty)}\right)^{-1}(z),&
            \qquad \mbox{for $z$ outside the disks,} \\[1ex]
        S(z)P^{-1}(z), & \qquad \mbox{for $z$ inside the disks.}
    \end{array}\right.
\end{equation}

\begin{remark}\label{Remark: invers exists}
    It is known that the inverses of the parametrices $P^{(\infty)}$ and $P$ exist,
    since all matrices have determinant one. For
    $P^{(\infty)}$, see (\ref{det-Pinfty=1}). For $P$ within the
    disks around the endpoints $a_i,b_j$ of $J$, as well as within the disks around the
    singular points of $V$ we refer to \cite{DKMVZ2}. For $P$ within the
    disk around the origin we refer to
    \cite[Section 4]{Vanlessen}.
\end{remark}

\begin{figure}
    \center{\resizebox{14cm}{!}{\includegraphics{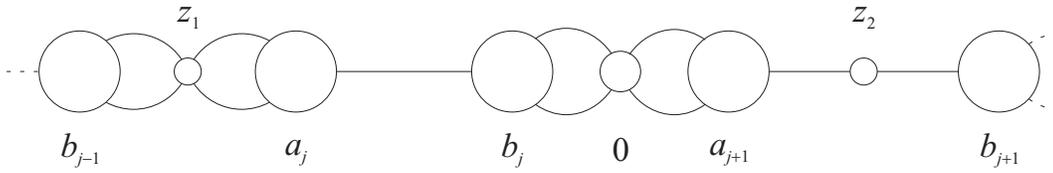}}
    \caption{Part of the contour $\Sigma_R$. The singular point $z_1$ corresponds to a point
    where $h$ vanishes at the interior of $J$, the singular point $z_2$ corresponds to a point
    where we obtain equality in (\ref{strict-equality-regular-case}).}\label{figure: system-of-contours-R}}
\end{figure}

Note that $P^{(\infty)}$ and $S$ have the same jumps on
$J\setminus\{0\}$, and that $P$ and $S$ have the same jumps on the
lens $\Sigma$ within the disks. This implies that $R$ is analytic
on the entire plane, except for jumps on the reduced system of
contours $\Sigma_R$, as shown in Figure \ref{figure:
system-of-contours-R}, cf.\ \cite{DKMVZ2}, and except for a
possible isolated singularity at the origin. Yet, as in
\cite{KMVV,Vanlessen}, it follows easily from the behavior of $S$
and $P$ near the origin, given by (\ref{RHPSd1}) and
(\ref{RHPSd2}), and by (\ref{RHPPd1}) and (\ref{RHPPd2}),
respectively,
that the isolated singularity of $R$ at the origin is removable.
Therefore $R$ is analytic on $\mathbb{C}\setminus\Sigma_R$.

Recall that the matrix valued functions $S$ and $P^{(\infty)}$ are
normalized at infinity. Since $\det P^{(\infty)}\equiv 1$, this
implies, by (\ref{R-in-function-of-S}), that also $R$ is
normalized at infinity.

Let $v_R$  be the jump matrix for $R$. It can be calculated
explicitly for each component of $\Sigma_R$. However, all that we
require are the following estimates, cf.\ \cite{DKMVZ2}
\begin{eqnarray}
    \nonumber
    \| v_R(z)\| &=&
        I+O(e^{-cn|z|}),\qquad \mbox{as $n\to\infty$, $z\in\Sigma_R\setminus$ circles,} \\[1ex]
    \nonumber
    \| v_R(z)\|&=&
        I+O(1/n^\kappa),\qquad \mbox{as $n\to\infty$, $z\in$ circles,}
\end{eqnarray}
for some $c>0$ and $0<\kappa\leq 1$, and where $\|\cdot \|$ is any
matrix norm. We note that the extra factor $|x|^{2\alpha}$, which
we will meet in $v_R$, does not cause any difficulties
to obtain this behavior.
These estimates then imply that $v_R$ is uniformly close to the
identity matrix as $n\to\infty$, and, since $R$ is normalized at
infinity, we then find uniformly for
$z\in\mathbb{R}\setminus\Sigma_R$,
\begin{equation}
    R(z)=I+O(1/n^\kappa),\qquad\mbox{as $n\to\infty$.}
\end{equation}
So, $R$ is uniformly bounded as $n\to\infty$. We also have that
$\frac{d}{dz}R(z)$ is uniformly bounded as $n\to\infty$. Another
useful property is $\det R\equiv 1$, which follows from
(\ref{R-in-function-of-S}) and the fact that $S, P^{(\infty)}$ and
$P$ all have determinant 1.


\section{Proof of Theorem \ref{Theorem: Universality at the origin}}
    \label{section: proof of the theorem}

We now have all the  ingredients necessary to prove Theorem
\ref{Theorem: Universality at the origin}. We point out that the
general scheme of this proof is the same as the proof of
\cite[Theorem 1.1(c)]{KV}. We replace in the kernel $K_n$, given
by (\ref{Kn}), the orthonormal polynomials $p_{n-1,n}$ and
$p_{n,n}$, together with their leading coefficients
$\gamma_{n-1,n}$ and $\gamma_{n,n}$, by the appropriate entries of
$Y$, given by (\ref{RHPYsolution}), and find
\begin{equation}\label{KninY}
    K_n(x,y)=-\frac{1}{2\pi
    i}\sqrt{w_n(x)}\sqrt{w_n(y)}\frac{Y_{11}(x)Y_{21}(y)-Y_{21}(x)Y_{11}(y)}{x-y}.
\end{equation}
This means that the kernel $K_n$ can be expressed in terms of the
first column of $Y$. Hence, we want to know the asymptotic
behavior of $Y$ near the origin. This will be determined in the
following lemma.

\begin{lemma}\label{Lemma: Behavior of Y near x0}
    For $x\in (0,\delta)$,
    \begin{equation}\label{Behavior of Y near x0}
        \begin{pmatrix}
            Y_{11}(x) \\
            Y_{21}(x)
        \end{pmatrix}
        =
        e^{-\frac{\pi i}{4}} \sqrt{\frac{\pi}{w_n(x)}} e^{\frac{n \ell}{2}\sigma_3} M_+(x)
        \begin{pmatrix}
            (n f(x))^{1/2} J_{\alpha+\frac{1}{2}}(n f(x))  \\[1ex]
            (n f(x))^{1/2} J_{\alpha-\frac{1}{2}}(n f(x))
        \end{pmatrix},
    \end{equation}
    with $M(z)$ given by
    \begin{equation}\label{M}
        M(z)=R(z) E(z)e^{n\phi_+(0)\sigma_3}e^{-\frac{\pi i}{4}
        \sigma_3}\frac{1}{\sqrt 2}
        \begin{pmatrix}
            1 & i \\
            i & 1
        \end{pmatrix},
    \end{equation}
    where $R$ is the result of the transformations $Y\to T\to S\to
    R$ of the RH problem, and the matrix valued function $E$ is
    given by
    {\rm (\ref{NTildeQuadrant1})}--{\rm (\ref{NTildeQuadrant4})}.
    The matrix valued function $M$ is analytic in $U_\delta$
    with $M(z)$ and $\frac{d}{dz}M(z)$ uniformly bounded
    for $z\in U_\delta$ as $n\to\infty$. Furthermore,
    \begin{equation} \label{determinantM=1}
        \det M(z) \equiv 1.
    \end{equation}
\end{lemma}

\begin{proof}
    We use the series of transformations $Y\to T\to S\to R$ and unfold them for
    $z$ inside the disk $U_\delta$ and in the right upper part of the lens,
    so that $z \in f^{-1}(I)$. Since
    $\omega(z)=z^{2\alpha}$ and $W(z)=z^{\alpha}e^{-\pi i\alpha}$ for our choice of $z$,
    see (\ref{definitie-omega}) and (\ref{definitie-W}),
    we have by (\ref{T-in-function-of-Y}), (\ref{S-in-function-of-T}),
    (\ref{definitie-P}) and (\ref{R-in-function-of-S})
    \begin{eqnarray}
        \nonumber
        Y(z) &=&
            e^{\frac{n\ell}{2}\sigma_3} R(z) E_n(z) \Psi_\alpha(n f(z))
            e^{-n\phi(z)\sigma_3}z^{-\alpha\sigma_3}e^{\pi i \alpha\sigma_3}
        \\[1ex]
        \label{Proof: Behavior of Y near x0: eq1}
        &&\qquad\times\,
            \begin{pmatrix}
                1 & 0 \\
                z^{-2\alpha}e^{-2n\phi(z)} & 1
            \end{pmatrix}e^{-\frac{n\ell}{2}\sigma_3}e^{ng(z)\sigma_3}.
    \end{eqnarray}
    We then get for the first column of $Y$,
    \begin{equation}\label{Proof: Behavior of Y near x0: eq2}
            \begin{pmatrix}
                Y_{11}(z) \\
                Y_{21}(z)
            \end{pmatrix}
            = z^{-\alpha}e^{n(g(z)-\phi(z)-\frac{\ell}{2})}
            e^{\frac{n \ell}{2}\sigma_3} R(z) E_n(z)
            \Psi_\alpha(nf(z))e^{\pi i \alpha \sigma_3}
            \begin{pmatrix}
                1 \\
                1
            \end{pmatrix}.
    \end{equation}
    Since $z$ is in the right upper part of the lens and inside
    the disk $U_\delta$, we have $0<\arg nf(z)<\pi/4$, cf.\
    Figure \ref{figure: conformal-mapping-f},
    and we thus use (\ref{PsiAlphaI}) to evaluate $\Psi_\alpha(nf(z))$.
    Using the formulas 9.1.3 and 9.1.4 of
    \cite{AbramowitzStegun} which connect the Hankel functions
    with the usual $J$-Bessel functions, we  find
    \begin{equation}\label{Proof: Behavior of Y near x0: eq3}
        \Psi_\alpha(nf(z))e^{\pi i \alpha\sigma_3}
            \begin{pmatrix}
                1 \\
                1
            \end{pmatrix}=
            e^{-\frac{\pi i}{4}}\sqrt\pi
            \begin{pmatrix}
                (n f(z))^{1/2} J_{\alpha+\frac{1}{2}}(n f(z)) \\[1ex]
                (n f(z))^{1/2} J_{\alpha-\frac{1}{2}}(n f(z))
            \end{pmatrix}.
    \end{equation}
    By (\ref{Enx0}) and (\ref{M}) we have
    $R(z)E_n(z)=M(z)$. Inserting this and
    (\ref{Proof: Behavior of Y near x0: eq3}) into (\ref{Proof: Behavior of Y near x0: eq2})
    we get
    \begin{equation}
        \begin{pmatrix}
            Y_{11}(z) \\
            Y_{21}(z)
        \end{pmatrix}=
        e^{-\frac{\pi i}{4}}\sqrt\pi z^{-\alpha}e^{n(g(z)-\phi(z)-\frac{\ell}{2})}
            e^{\frac{n\ell}{2}\sigma_3} M(z)
        \begin{pmatrix}
            (n f(z))^{1/2} J_{\alpha+\frac{1}{2}}(n f(z)) \\[1ex]
            (n f(z))^{1/2} J_{\alpha-\frac{1}{2}}(n f(z))
        \end{pmatrix}.
    \end{equation}
    Letting $z\to x\in(0,\delta)$, and noting that
    \begin{equation}\label{Proof: Behavior of Y near x0: eq4}
        x^{-\alpha}e^{n(g_+(x)-\phi_+(x)-\frac{\ell}{2})} =
            x^{-\alpha}e^{\frac{1}{2}n(g_+(x)+g_-(x)-\ell)} =
            x^{-\alpha}e^{\frac{1}{2}n V(x)} =
            w_n(x)^{-1/2},
    \end{equation}
    which follows from the fact that $2\phi_+(x)=g_+(x)-g_-(x)$, see
    Section \ref{section: second transformation}, and from
    (\ref{varying-weight}) and (\ref{crucial-property-g-1}),
    we obtain (\ref{Behavior of Y near x0}).

    The matrix valued function $M$ is analytic in the disk
    $U_\delta$ since both $R$ and $E$ are analytic
    in this disk. So, we may write $M(x)$ instead of
    $M_+(x)$ in (\ref{Behavior of Y near x0}).

    We recall that
    $R(z)$, $\frac{d}{dz}R(z)$, $E(z)$ and $\frac{d}{dz}E(z)$ are uniformly bounded for $z\in
    U_\delta$ as $n\to\infty$, see Section \ref{section: parametrix near the origin} and
    Section \ref{section: third transformation}. If we also
    use that $|e^{n\phi_+(0)}|=1$, which follows from the fact
    that $\phi_+$ is purely imaginary on $J$, we have from (\ref{M}) that $M(z)$
    and $\frac{d}{dz}M(z)$ are uniformly bounded for $z\in
    U_\delta$ as $n\to\infty$.

    Since $M$ is a product of five matrices all with determinant
    one, (\ref{determinantM=1}) is true.
\end{proof}

\begin{lemma}\label{Lemma: Asymptotics of notations}
    Let $u\in (0,\infty)$, $u_n=\frac{u}{n\psi(0)}$ and $\tilde
    u_n=nf(u_n)$. Then
    \begin{equation}\label{universalitylemmautilden}
        \tilde u_n=\pi u+O\left(\frac{u^2}{n}\right),\qquad
        \mbox{as $n\to\infty$,}
    \end{equation}
    \begin{equation}\label{universalitylemmaJ+}
        J_{\alpha+\frac{1}{2}}(\tilde
        u_n)=J_{\alpha+\frac{1}{2}}(\pi
        u)+O\left(\frac{u^{\alpha+\frac{3}{2}}}{n}\right),
        \qquad\mbox{as $n\to\infty$,}
    \end{equation}
    \begin{equation}\label{universalitylemmaJ-}
        J_{\alpha-\frac{1}{2}}(\tilde
        u_n)=J_{\alpha-\frac{1}{2}}(\pi
        u)+O\left(\frac{u^{\alpha+\frac{1}{2}}}{n}\right),
        \qquad\mbox{as $n\to\infty$,}
    \end{equation}
    where the error terms hold uniformly for $u$ in bounded
    subsets of $(0,\infty)$.
\end{lemma}

\begin{proof}
    Since, see (\ref{behavior-f})
    \[
        f(x)=\pi\psi(0)x+O(x^2),
        \qquad\mbox{as $x\to 0$,}
    \]
    we have, uniformly for $u$ in bounded subsets of $(0,\infty)$,
    \[
        f\left(\frac{u}{n\psi(0)}\right)=\pi\frac{u}{n}
            +O\left(\frac{u^2}{n^2}\right), \qquad\mbox{as $n\to\infty$,}
    \]
    which proves (\ref{universalitylemmautilden}).

    We note \cite[formula 9.1.10]{AbramowitzStegun} that
    $J_{\alpha+\frac{1}{2}}(z)=z^{\alpha+\frac{1}{2}}H(z)$, with
    $H$ an entire function. It then follows from
    (\ref{universalitylemmautilden}) that, as
    $n\to\infty$, uniformly for $u$ in bounded subsets of
    $(0,\infty)$,
    \begin{eqnarray}
        \nonumber
        J_{\alpha+\frac{1}{2}}(\tilde u_n)
        &=&
            \left[(\pi u)^{\alpha+\frac{1}{2}}+O\left(\frac{u^{\alpha+\frac{3}{2}}}{n}\right)\right]
            \left[H(\pi u)+O\left(\frac{u^2}{n}\right)\right] \\[2ex]
        \nonumber
        &=&
            J_{\alpha+\frac{1}{2}}(\pi u)+O\left(\frac{u^{\alpha+\frac{3}{2}}}{n}\right),
    \end{eqnarray}
    so that equation (\ref{universalitylemmaJ+}) is proved.
    Similarly, we can prove (\ref{universalitylemmaJ-}).
\end{proof}

We are now able to prove Theorem \ref{Theorem: Universality at the
origin}.

\begin{varproof}\textbf{of Theorem \ref{Theorem: Universality at the origin}.}
    Let $u,v\in (0,\infty)$ and define
    \[
        u_n=\frac{u}{n\psi(0)},\quad
        v_n=\frac{v}{n\psi(0)},\quad
        \tilde{u}_n=nf(u_n),\quad
        \tilde{v}_n=nf(v_n).
    \]
    We put
    \[
        \hat K_n(u,v)=\frac{1}{n\psi(0)}K_n(u_n,v_n).
    \]
    From (\ref{KninY}) and (\ref{Behavior of Y near x0}) we then have
    \begin{eqnarray}
        \nonumber
            \hat K_n(u,v)
        &=&
            -\frac{1}{2\pi i(u-v)} \det
            \begin{pmatrix}
                e^{-\frac{n\ell}{2}} \sqrt{w_n(u_n)} Y_{11}(u_n) & e^{-\frac{n\ell}{2}} \sqrt{w_n(v_n)} Y_{11}(v_n) \\[1ex]
                e^{\frac{n\ell}{2}} \sqrt{w_n(u_n)} Y_{21}(u_n) &  e^{\frac{n\ell}{2}} \sqrt{w_n(v_n)} Y_{21}(v_n)
            \end{pmatrix} \\[2ex]
        \nonumber
        &=&
            \frac{1}{2 (u-v)} \det
            \left[
            M(u_n)
            \begin{pmatrix}
                \tilde{u}_n^{1/2} J_{\alpha+\frac{1}{2}}(\tilde{u}_n) & 0 \\[1ex]
                \tilde{u}_n^{1/2} J_{\alpha-\frac{1}{2}}(\tilde{u}_n) & 0
            \end{pmatrix}
            + M(v_n)
            \begin{pmatrix}
                0 & \tilde{v}_n^{1/2} J_{\alpha+\frac{1}{2}}(\tilde{v}_n) \\[1ex]
                0 & \tilde{v}_n^{1/2} J_{\alpha-\frac{1}{2}}(\tilde{v}_n)
            \end{pmatrix}\right].
    \end{eqnarray}
    The matrix in the determinant can be written as
    \begin{eqnarray}
        \nonumber
        \lefteqn{
            M(v_n)\left[
            \begin{pmatrix}
                \tilde{u}_n^{1/2} J_{\alpha+\frac{1}{2}}(\tilde{u}_n) &
                    \tilde{v}_n^{1/2} J_{\alpha+\frac{1}{2}}(\tilde{v}_n) \\[1ex]
                \tilde{u}_n^{1/2} J_{\alpha-\frac{1}{2}}(\tilde{u}_n) &
                    \tilde{v}_n^{1/2} J_{\alpha-\frac{1}{2}}(\tilde{v}_n)
            \end{pmatrix}\right.
        } \\[2ex]
        \label{Proof: Universality: eq1}
        & &
            \qquad \left.
            +\, M(v_n)^{-1}(M(u_n)-M(v_n))
            \begin{pmatrix}
                \tilde{u}_n^{1/2} J_{\alpha+\frac{1}{2}}(\tilde{u}_n) & 0 \\[1ex]
                \tilde{u}_n^{1/2} J_{\alpha-\frac{1}{2}}(\tilde{u}_n) & 0
            \end{pmatrix}
            \right].
    \end{eqnarray}
    We will now determine the asymptotics of the second
    term in (\ref{Proof: Universality: eq1}). Since $\det M(v_n)=1$
    and since $M(z)$ is uniformly bounded for $z\in
    U_\delta$, see Lemma \ref{Lemma: Behavior of Y near x0}, the
    entries of $M(v_n)^{-1}$ are uniformly bounded. By Lemma
    \ref{Lemma: Behavior of Y near x0} we also have that $\frac{d}{dz}M(z)$
    is uniformly bounded for $z\in U_\delta$, so that from
    the mean value theorem
    $M(u_n)-M(v_n)=O\left(\frac{u-v}{n}\right)$. From Lemma
    \ref{Lemma: Asymptotics of notations} it follows that $\tilde u_n^{1/2}J_{\alpha+\frac{1}{2}}(\tilde
    u_n)=O(u^{\alpha+1})$ and $\tilde u_n^{1/2}J_{\alpha-\frac{1}{2}}(\tilde
    u_n)=O(u^\alpha)$ uniformly for $u$ in bounded subsets of
    $(0,\infty)$ as $n\to\infty$. Hence we have, uniformly for
    $u,v$ in bounded subsets of $(0,\infty)$,
    \[
        M(v_n)^{-1}(M(u_n)-M(v_n))
        \begin{pmatrix}
                \tilde{u}_n^{1/2} J_{\alpha+\frac{1}{2}}(\tilde{u}_n) & 0 \\[1ex]
                \tilde{u}_n^{1/2} J_{\alpha-\frac{1}{2}}(\tilde{u}_n) & 0
            \end{pmatrix}
        =
        \begin{pmatrix}
            O\left(\frac{u-v}{n}u^\alpha\right) & 0 \\[1ex]
            O\left(\frac{u-v}{n}u^\alpha\right) & 0
        \end{pmatrix}.
    \]
    Inserting this into (\ref{Proof: Universality: eq1}), using the fact that $\det
    M(v_n)=1$, and that $\tilde v_n^{1/2}J_{\alpha\pm\frac{1}{2}}(\tilde
    v_n)=O(v^\alpha)$ as $n\to\infty$, we then find uniformly for
    $u,v$ in bounded subsets of $(0,\infty)$,
    \begin{eqnarray}
        \nonumber
            \hat K_n(u,v)
        &=&
            \frac{1}{2(u-v)}\det
            \begin{pmatrix}
                \tilde{u}_n^{1/2} J_{\alpha+\frac{1}{2}}(\tilde{u}_n)+O\left(\frac{u-v}{n}u^\alpha\right) &
                    \tilde{v}_n^{1/2} J_{\alpha+\frac{1}{2}}(\tilde{v}_n) \\[1ex]
                \tilde{u}_n^{1/2} J_{\alpha-\frac{1}{2}}(\tilde{u}_n)+O\left(\frac{u-v}{n}u^\alpha\right) &
                    \tilde{v}_n^{1/2} J_{\alpha-\frac{1}{2}}(\tilde{v}_n)
            \end{pmatrix} \\[2ex]
        &=&
            \frac{1}{2(u-v)}\det
            \begin{pmatrix}
                \tilde{u}_n^{1/2} J_{\alpha+\frac{1}{2}}(\tilde{u}_n) &
                    \tilde{v}_n^{1/2} J_{\alpha+\frac{1}{2}}(\tilde{v}_n) \\[1ex]
                \tilde{u}_n^{1/2} J_{\alpha-\frac{1}{2}}(\tilde{u}_n) &
                    \tilde{v}_n^{1/2} J_{\alpha-\frac{1}{2}}(\tilde{v}_n)
            \end{pmatrix}
            +O\left(\frac{u^\alpha v^\alpha}{n}\right).
    \end{eqnarray}
    We note, from Lemma \ref{Lemma: Asymptotics of notations},
    that we can replace in the determinant, $\tilde u_n$ by $\pi u$ and $\tilde v_n$ by $\pi
    v$. We then make an error which can be estimated by Lemma
    \ref{Lemma: Asymptotics of notations}. However, since this estimate is
    not uniform for $u-v$ close to zero, we have to be more
    careful. We insert a factor $u^{-\alpha}$ in the first column of the determinant, and
    a factor $v^{-\alpha}$ in the second. Then we subtract the second
    column from the first to obtain
    \begin{eqnarray}
        \nonumber
            \hat K_n(u,v)& = &\frac{u^{\alpha}v^{\alpha}}{2(u-v)}
            \det
            \begin{pmatrix}
                u^{-\alpha}\tilde{u}_n^{1/2} J_{\alpha+\frac{1}{2}}(\tilde{u}_n) -
                    v^{-\alpha}\tilde{v}_n^{1/2} J_{\alpha+\frac{1}{2}}(\tilde{v}_n) &
                    v^{-\alpha}\tilde{v}_n^{1/2} J_{\alpha+\frac{1}{2}}(\tilde{v}_n) \\[1ex]
                u^{-\alpha}\tilde{u}_n^{1/2} J_{\alpha-\frac{1}{2}}(\tilde{u}_n) -
                    v^{-\alpha}\tilde{v}_n^{1/2} J_{\alpha-\frac{1}{2}}(\tilde{v}_n) &
                    v^{-\alpha}\tilde{v}_n^{1/2} J_{\alpha-\frac{1}{2}}(\tilde{v}_n)
            \end{pmatrix} \\[1ex]
        \label{Proof: Universality: eq2}
        &&
            \qquad +\,
            O\left(\frac{u^\alpha v^\alpha}{n}\right).
    \end{eqnarray}
    One can check, using (\ref{universalitylemmautilden}),
    (\ref{universalitylemmaJ+})
    and the facts that $J'_{\alpha+\frac{1}{2}}(\tilde x_n)=J'_{\alpha+\frac{1}{2}}(\pi x)+
    O(\frac{x^{\alpha+\frac{1}{2}}}{n})$ and $\frac{d}{dx}\tilde x_n=\pi+O(\frac{x}{n})$,
    where we have put $\tilde{x}_n = n f(\frac{x}{n\psi(0)})$,
    that
    \[
        \frac{d}{dx}\left[
        x^{-\alpha}\tilde x_n^{1/2}J_{\alpha+\frac{1}{2}}(\tilde
        x_n)-x^{-\alpha}(\pi x)^{1/2}J_{\alpha+\frac{1}{2}}(\pi x)
        \right]
        =O\left(\frac{x}{n}\right), \qquad \mbox{as $n\to\infty$,}
    \]
    uniformly for $x$ in bounded subsets of $(0,\infty)$.
    It then follows that the (1,1)--entry in the determinant of (\ref{Proof: Universality: eq2})
    is equal to
    \[
        u^{-\alpha}(\pi u)^{1/2}J_{\alpha+\frac{1}{2}}(\pi u)
                    -v^{-\alpha}(\pi v)^{1/2}J_{\alpha+\frac{1}{2}}(\pi
                    v)+O\left(\frac{u-v}{n}\right).
    \]
    Similarly, we have from
    \[
        \frac{d}{dx}\left[
        x^{-\alpha}\tilde x_n^{1/2}J_{\alpha-\frac{1}{2}}(\tilde
        x_n)-x^{-\alpha}(\pi x)^{1/2}J_{\alpha-\frac{1}{2}}(\pi x)
        \right]
        =O\left(\frac{1}{n}\right), \qquad \mbox{as $n\to\infty$,}
    \]
    that the (2,1)--entry in the determinant of (\ref{Proof: Universality: eq2})
    is equal to
    \[
        u^{-\alpha}(\pi u)^{1/2}J_{\alpha-\frac{1}{2}}(\pi u)
                    -v^{-\alpha}(\pi v)^{1/2}J_{\alpha-\frac{1}{2}}(\pi
                    v)+O\left(\frac{u-v}{n}\right).
    \]
    From Lemma \ref{Lemma: Asymptotics of notations} it also
    follows that $\tilde v_n J_{\alpha\pm\frac{1}{2}}(\tilde
    v_n)=(\pi v)J_{\alpha\pm\frac{1}{2}}(\pi v)+O(1/n)$.
    Therefore, uniformly for $u,v$ in bounded subsets of
    $(0,\infty)$,
    \begin{eqnarray}
        \nonumber
        \lefteqn{
            \hat K_n(u,v) = \frac{u^\alpha v^\alpha}{2(u-v)}
        } \\[2ex]
        \nonumber
        & &
            \times\, \det
            \begin{pmatrix}
                u^{-\alpha}(\pi u)^{1/2}J_{\alpha+\frac{1}{2}}(\pi u)
                    -v^{-\alpha}(\pi v)^{1/2}J_{\alpha+\frac{1}{2}}(\pi v)+O\left(\frac{u-v}{n}\right) &
                    v^{-\alpha}(\pi v)^{1/2}J_{\alpha+\frac{1}{2}}(\pi v) + O\left(\frac{1}{n}\right) \\[1ex]
                u^{-\alpha}(\pi u)^{1/2}J_{\alpha-\frac{1}{2}}(\pi u)
                    -v^{-\alpha}(\pi v)^{1/2}J_{\alpha-\frac{1}{2}}(\pi v)+O\left(\frac{u-v}{n}\right) &
                    v^{-\alpha}(\pi v)^{1/2}J_{\alpha-\frac{1}{2}}(\pi v) + O\left(\frac{1}{n}\right)
            \end{pmatrix} \\[2ex]
        \nonumber
        & &
            \qquad +\, O\left(\frac{u^\alpha v^\alpha}{n}\right) \\[2ex]
        \nonumber
        &=&
            \mathbb{J}_{\alpha}^o(u,v)
            + \frac{u^\alpha v^\alpha}{2(u-v)} \det
            \begin{pmatrix}
                u^{-\alpha}(\pi u)^{1/2}J_{\alpha+\frac{1}{2}}(\pi u)
                    -v^{-\alpha}(\pi v)^{1/2}J_{\alpha+\frac{1}{2}}(\pi v) &
                    O\left(\frac{1}{n}\right) \\[1ex]
                u^{-\alpha}(\pi u)^{1/2}J_{\alpha-\frac{1}{2}}(\pi u)
                    -v^{-\alpha}(\pi v)^{1/2}J_{\alpha-\frac{1}{2}}(\pi v) &
                    O\left(\frac{1}{n}\right)
            \end{pmatrix} \\[1ex]
        \label{Proof: Universality: eq3}
        & &
            \qquad
            +\, O\left(\frac{u^\alpha v^\alpha}{n}\right)
    \end{eqnarray}
    Since $z^{-\alpha+\frac{1}{2}}J_{\alpha\pm\frac{1}{2}}(z)$
    is an entire function we have by the mean value
    theorem that
    \[
        \frac{u^{-\alpha}(\pi u)^{1/2}J_{\alpha\pm\frac{1}{2}}(\pi u)
                    -v^{-\alpha}(\pi v)^{1/2}J_{\alpha\pm\frac{1}{2}}(\pi v)}{u-v}
    \]
    is bounded for $u,v$ in bounded subsets of $(0,\infty)$.
    From (\ref{Proof: Universality: eq3}) we then have
    \[
        \hat K_n(u,v)=\mathbb{J}_{\alpha}^o(u,v)+O\left(\frac{u^\alpha
        v^\alpha}{n}\right),
    \]
    uniformly for $u,v$ in bounded subsets of $(0,\infty)$, which
    completes the proof of Theorem \ref{Theorem: Universality at the origin}.
\end{varproof}


\end{document}